\documentstyle[preprint,eqsecnum,aps,epsf,rotate,floats]{revtex}

\begin{document}

%\tighten

\renewcommand{\thefootnote}{\fnsymbol{footnote}}
\begin{titlepage}
\begin{flushright}
CERN-TH/98-240\\
\mbox{CMU-9805\hspace*{0.6cm}}\\
hep-ph/9808360\\
\mbox{August 1998\hspace*{0.45cm}}\\
\end{flushright}
\vskip 1.2cm
\begin{center}
 \boldmath
{\Large\bf QCD analysis of inclusive {\em B} decay \\[0.2cm]
into charmonium}
\unboldmath
%---------------------------------------------------------
\vskip 1.8cm
{\sc M. Beneke} \hspace{0.2cm}and \hspace{0.2cm} 
{\sc F. Maltoni}\footnote{Permanent address: 
Dipartimento di Fisica dell'Universit\`a 
and Sez. INFN, Pisa, Italy}
\vskip .3cm
{\it Theory Division, CERN, CH-1211 Geneva 23}
%---------------------------------------------------------
\vskip .4cm
and
\vskip .4cm
{\sc I.Z. Rothstein}
\vskip .3cm
{\em Department of Physics, Carnegie Mellon University,
 
Pittsburgh, PA 15213, U.S.A}
%---------------------------------------------------------
\vskip 0.0cm
\end{center}

\begin{abstract}
\noindent\hspace*{-0.34cm} We compute 
the decay rates and $H$-energy distributions of $B$ mesons  
into the final state $H+X$, where $H$ can be any one of the 
$S$-wave or $P$-wave charmonia, at next-to-leading order in the 
strong coupling. We find that a significant fraction of the 
observed $J/\psi$, $\psi'$ and $\chi_c$ must be produced 
through $c\bar{c}$ pairs in a colour octet state and should 
therefore be accompanied by more than one light hadron. At the 
same time we obtain stringent constraints on some of the 
long-distance parameters for colour octet production.\\  

\noindent PACS Nos.: 13.25.Hw, 14.40.Gx, 12.38.Bx
\end{abstract}

\vfill

\end{titlepage}

\section{Introduction}

\noindent
Exclusive $B$ decays provide us with important information on the 
structure of the Cabibbo-Kobayashi-Maskawa (CKM) matrix. However,  
the theoretical calculation of absolute branching fractions is 
complicated by the fact that a rather detailed knowledge of strong 
interaction effects is required. The theoretical situation with 
regard to strong interaction dynamics 
improves as one considers more inclusive final states. At leading order  
in $\Lambda_{QCD}/m_b$, where $m_b$ is the $b$ quark mass and 
$\Lambda_{QCD}$ the strong interaction scale, 
the totally inclusive $B$ decay rate 
can be computed completely in perturbation theory. However, it is not 
necessary that the process be totally inclusive. A semi-inclusive 
decay $B\to H+X$ can also be treated perturbatively in part, provided 
the formation of the hadron $H$ proceeds through a short-distance process. 
This is the case, if $H$ is a charmonium state, because the production 
of a charm-quark $c\bar{c}$ pair requires energies much larger than 
$\Lambda$. The bound state dynamics of $H$ then factorizes and can be 
parametrized. This statement is summarized by the factorization formula 
\cite{BBL95}
\begin{equation}
\label{factform}
\Gamma(B\to H+X) = \sum_n C(b\to c\bar{c}[n]+x)\,\langle
{\cal O}^H[n]\rangle,
\end{equation}
which is valid up to power corrections of order $\Lambda_{QCD}/m_{b,c}$. 
(To this 
accuracy it is justified to treat the $B$ meson as a free $b$ quark.) 
The parameters $\langle{\cal O}^H[n]\rangle$, defined in \cite{BBL95}, 
are sensitive to the charmonium 
bound state scales of order $m_c v$ and $m_c v^2$, where $v$ is the 
typical charm quark velocity in the charmonium bound state. With 
$v^2\sim 0.25$ for $J/\psi$ we consider these scales to be too small to be 
treated perturbatively. On the other hand the coefficient functions 
$C(b\to c\bar{c}[n]+x)$ describe the production of a $c\bar{c}$ 
configuration $n$ at short distances and can be expanded in the 
strong coupling $\alpha_s(\mu)$ at a scale $\mu$ of order $2 m_c$.
We have expressed the decay rate in terms of several non-perturbative 
parameters $\langle {\cal O}^H[n]\rangle$. The predictive power 
lies in the fact that these 
parameters are independent of the particular charmonium production 
process and hence 
are constrained by other charmonium production processes. 

Because charmonia pass as non-relativistic systems, Eq.~(\ref{factform}) 
involves an expansion in $v^2$ and the $c\bar{c}[n]$ configurations 
that appear in lower orders of this expansion can be usefully classified 
by ${}^{2S+1}\!L_J^{(C)}$, where $S$, $L$ and $J$ refer to spin, 
orbital angular momentum and total angular momentum, respectively. 
In addition $C=1,8$ refers to a colour singlet or a colour octet 
configuration. In the present work we calculate the short-distance 
coefficients for 
\begin{equation}
\label{set}
n\in\left\{{}^3\!S_1^{(1,8)},{}^1\!S_0^{(1,8)},
{}^3\!P_{0,1,2}^{(1)},{}^3\!P_{J}^{(8)},{}^1\!P_1^{(1,8)}\right\}
\end{equation}
at next-to-leading order (NLO) in $\alpha_s$. As we discuss later, 
we believe that these terms in the velocity expansion 
are sufficient to reliably predict 
(to about 25\%, barring radiative corrections in $\alpha_s$) the 
decay rates into $J/\psi$, $\psi'$, $\eta_c$, 
$\chi_{0,1,2}$ and the elusive ${}^1P_1$ state $h_c$ as a function of the 
long-distance parameters $\langle {\cal O}^H[n]\rangle$. We find that, 
given the present uncertainties in the long-distance parameters, 
the experimentally observed branching fraction \cite{CLEO95} for 
$J/\psi$ and $\psi'$ can 
easily be accounted for at NLO. However, we find it difficult to account 
for the observed $\chi_{c1}$ and $\chi_{c2}$ branching fractions 
simultaneously, because the $\alpha_s$ expansion of the colour 
singlet contribution in $\chi_{c1}$ production turns out to be untrustworthy 
at NLO. The NLO corrections typically enhance 
the decay rate by about (20-50)\% in the colour octet channels and lead to 
bounds on the long-distance parameters, which should be useful for the 
phenomenology of other charmonium production processes. We also compute 
weights of the charmonium energy distribution, which yield additional 
information. The shape of the energy distribution itself, however, 
is difficult to predict, because it is distorted by the motion of the 
$b$ quark in the $B$ meson and the energy taken away in the hadronization 
of a $c\bar{c}$ state $n$. This distortion averages out in weighted 
sums, as long as the weights are sufficiently smooth. 

We then 
compare the inclusive calculation for $J/\psi+X$ with the sum of the 
measured decay rates for $J/\psi+K$ and  $J/\psi+K^*$. 
The comparison suggests a significant fraction of 
multi-body decays, consistent with the energy spectrum observed by 
CLEO \cite{CLEO95}. A substantial contribution from multi-body decays 
is also reassuring from the point of view of validity of the theoretical 
calculation. Factorization implies that a $c\bar{c}$ state $n$ 
hadronizes into a $J/\psi$ plus light hadrons independent of the 
remaining decay process up to corrections of order $\Lambda_{QCD}/m_{b,c}$. 
If $n$ refers to a colour octet state, 
the conversion into charmonium requires the emission of at least 
one gluon. Although colour reconnections with the spectator 
quark in the $B$ meson must eventually occur, we expect a charmonium 
produced through a colour octet $c\bar{c}$ state to be 
accompanied by more than one light hadron more often than 
for a colour singlet $c\bar{c}$ state. Since we find that a 
large fraction of the total decay rate is from colour octet 
intermediate states, we also expect a large fraction of multi-body final 
states. This evidence also suggests to us that the energy released 
in the $B$ meson decay into charmonium is already large enough for 
an inclusive treatment to be applicable.

Inclusive production of $S$ wave charmonia has been considered in 
Refs.~\cite{DT80,Wise80} in the colour singlet model and at leading 
order (LO). In addition, the colour singlet 
production of the $P$-wave state 
$\chi_{c1}$ was computed in Ref.~\cite{KNR80}. (At LO the states 
$\chi_{c0}$ and $\chi_{c2}$ are not produced.) 
The colour singlet model is contained 
in Eq.~(\ref{factform}) as the term where the quantum numbers of 
$n$ match those of the charmonium state. For $P$-wave charmonia 
the colour singlet model does not coincide with the non-relativistic 
limit $v\to 0$ and is generally inconsistent. The authors of 
Ref.~\cite{BBYL92} noted that the contribution from 
$c\bar{c}[{}^3\!S_1^{(8)}]$ is leading order in $v$ for $\chi_{cJ}$ and that 
$c\bar{c}[{}^1\!S_0^{(8)}]$ is leading order for $h_c$. They 
computed the relevant short-distance coefficients to LO in 
$\alpha_s$. In the case of $J/\psi$, the short-distance coefficients 
of $c\bar{c}[n]$ states 
with $n={}^3S_1^{(8)},{}^1\!S_0^{(8)},{}^3P_J^{(8)}$ are strongly 
enhanced as a consequence of the particular structure of the 
weak effective Lagrangian that mediates $b$ quark decay. These 
production channels have to be taken into account although the 
corresponding long-distance matrix elements are suppressed by a 
factor of $v^4$. The relevant coefficient functions were computed 
in Ref.~\cite{KLS96}, again at LO in $\alpha_s$. Ref.~\cite{FHMN97} 
adds a study of $J/\psi$ polarization effects. The only NLO 
calculation of charmonium production in $B$ decay is due to 
Bergstr\"om and Ernstr\"om \cite{BE94}, who computed the contribution 
of the colour singlet ${}^3\!S_1$ intermediate $c\bar{c}$ state 
to $J/\psi$ production. We repeated their calculation and comment on it 
later on.

The paper is organized as follows: In Section~\ref{notation} we 
introduce notation and discuss the structure of important contributions 
to a given charmonium state. Section~\ref{calculation} provides some 
details on the calculation related to the handling of ultraviolet and 
infrared divergences at intermediate stages. Section~\ref{results} 
contains our main results. We present expressions for the decay rates 
in numerical form and a comparison with existing experimental data. 
Analytic results for the decay rates and energy distributions are 
collected in two appendices for reference. Section~\ref{conclusion} 
contains our conclusions.

\section{Preliminaries}
\label{notation}

\noindent
The terms of interest in the $\Delta B = 1$ effective weak Hamiltonian  
\begin{equation}  
H_{eff} = \frac{G_{F}}{\sqrt{2}} \sum_{q=s,d} 
\left\{V_{cb}^\ast V_{cq} \,
\left[ \frac{1}{3} C_{[1]}(\mu) {\cal O}_1(\mu) + C_{[8]}(\mu) 
{\cal O}_8(\mu) \right] - V_{tb}^\ast V_{tq} \sum_{i=3}^6 
C_i(\mu) {\cal O}_i(\mu)\right\}
\label{eq:Heff}  
\end{equation}
contain the `current-current' operators   
\begin{eqnarray}
{\cal O}_1 &=&   
[\bar{c} \gamma_{\mu} (1-\gamma_5) c] \, 
[\bar{b} \gamma^{\mu} (1-\gamma_5) q]
\label{eq:opsing} \\  
{\cal O}_8 &=&    
[\bar{c}\,T^A \gamma_{\mu} (1-\gamma_5) c]\,  
[\bar{b}\,T^A \gamma^{\mu} (1-\gamma_5) q] 
\label{eq:opoct}  
\end{eqnarray}
and the QCD penguin operators ${\cal O}_{3-6}$. (See the review 
Ref.~\cite{BBL96} for 
their precise definition.) For the decays $B\to \mbox{charmonium}\,+X$ 
it is convenient to choose a Fierz version of the current-current 
operators such that the $c\bar{c}$ pair at the weak decay vertex is 
either in a colour singlet or a colour octet state. The coefficient 
functions are related to the usual $C_\pm$ by  
\begin{eqnarray}
C_{[1]}(\mu) &=& 2 C_+(\mu) - C_-(\mu),  
\label{eq:c1nlo} \\  
C_{[8]}(\mu) &=&   C_+(\mu) + C_-(\mu).  
\label{eq:c8nlo}  
\end{eqnarray}
The NLO Wilson coefficients $C_\pm(\mu)$ have been computed in 
Refs.~\cite{ACMP81,BW90}. With the conventions of Ref.~\cite{BW90}   
\begin{equation}
C_\pm(\mu) = \left[ \frac{\alpha_s(M_W)}{\alpha_s(\mu)} \right]  
^{\gamma^{(0)}_\pm/(2 \beta_0)}
\left( 1 + \frac{\alpha_s(\mu)}{4\pi} B_\pm  \right) 
\left( 1 + \frac{\alpha_s(M_W)-\alpha_s(\mu)}{4\pi} (B_\pm-J_\pm) \right)
\label{eq:Cpmnlo}   
\end{equation} 
with  
\begin{eqnarray}
J_\pm &=&\frac{\gamma^{(0)}_\pm\beta_1}{2 \beta_0^2} -   
\frac{\gamma^{(1)}_\pm}{2 \beta_0}\\
B_\pm &=& \frac{3\mp 1}{6} \,(\pm 11 + \kappa_\pm)
\end{eqnarray}
and the one-loop and two-loop anomalous dimensions
\begin{eqnarray}
\gamma^{(0)}_\pm &=& \pm \,2 \,(3 \mp 1),\\ 
\gamma^{(1)}_\pm &=& \frac{3 \mp 1}{6} \left(-21\pm\frac{4}{3}\,n_f   
- 2 \beta_0 \kappa_\pm \right).
\end{eqnarray}
The quantity $\kappa_\pm$ is scheme-dependent and depends in particular 
on the treatment of $\gamma_5$. In the `naive dimensional 
regularization' (NDR) scheme, $\kappa_\pm= 0$; in the 't Hooft-Veltman 
(HV) scheme, $\kappa_\pm= \mp 4$. In the HV scheme the current-current 
operators, implied by the convention used in Refs.~\cite{BBL96,BW90}, are 
not minimally subtracted. If one computes the low energy matrix elements 
of the weak Hamiltonian in the modified minimal subtraction 
($\overline{\rm MS}$) scheme, as we will do below, one has to apply 
an additional finite renormalization. This amounts to multiplying the 
coefficients $C_\pm(\mu)$ by a factor of $1-4\alpha_s(\mu)/(3 \pi)$, 
or, equivalently, to an additional contribution to $\kappa_\pm$ in the 
HV scheme. No 
additional renormalization is required in the NDR scheme. 
At NLO the strong coupling is given by 
\begin{equation}
\alpha_s(\mu) = \frac{4 \pi}{\beta_0 \ln (\mu^2/\Lambda_{QCD}^2)}  
\left[ 1 - \frac{\beta_1 \ln [\ln (\mu^2/\Lambda_{QCD}^2)]}  
{\beta_0^2 \ln (\mu^2/\Lambda_{QCD}^2)} \right]    
\label{eq:asnlo}  
\end{equation}
with 
\begin{equation}
\beta_0 = 11 - \frac{2}{3} \, n_f, \qquad\quad
\beta_1 = 102 - \frac{38}{3} \, n_f. 
\end{equation}

%%%%%%%%%%%%%%%%%%%%%%%%%%%%%%%%%%%%%%%%%%%%%%%%%%%%%%%%%%%%%%%%%%%
\begin{figure}[t]
   \vspace{-3.3cm}
   \epsfysize=23cm
   \epsfxsize=15.6cm
   \centerline{\epsffile{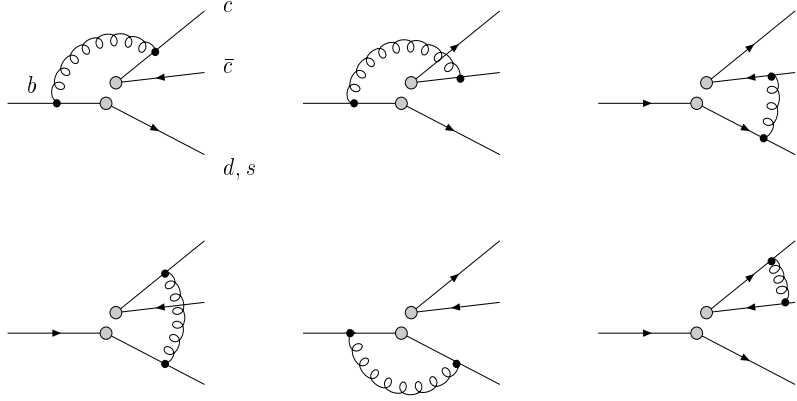}}
   \vspace*{-13.2cm}
\caption[dummy]{One-loop virtual corrections to 
$b\to c\bar{c} q$. Wave function renormalizations are 
not shown.\label{figa}}
\end{figure}
%%%%%%%%%%%%%%%%%%%%%%%%%%%%%%%%%%%%%%%%%%%%%%%%%%%%%%%%%%%%%%%%%%%
%%%%%%%%%%%%%%%%%%%%%%%%%%%%%%%%%%%%%%%%%%%%%%%%%%%%%%%%%%%%%%%%%%%
\begin{figure}[t]
   \vspace{-3cm}
   \epsfysize=23cm
   \epsfxsize=15.6cm
   \centerline{\epsffile{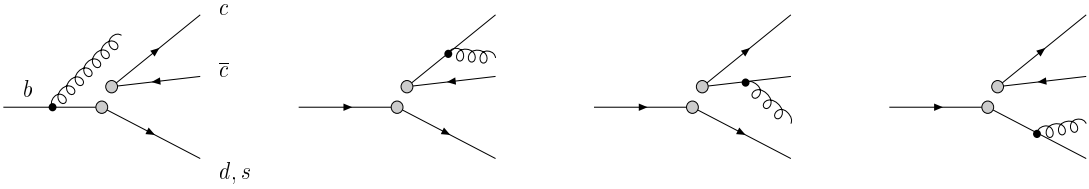}}
   \vspace*{-16.2cm}
\caption[dummy]{Real gluon corrections to $b\to c\bar{c} q$.
\label{figb}}
\end{figure}
%%%%%%%%%%%%%%%%%%%%%%%%%%%%%%%%%%%%%%%%%%%%%%%%%%%%%%%%%%%%%%%%%%%
The NLO QCD corrections involve the one-loop virtual gluon correction 
to $b\to c\bar{c}[n]+q$ and the real gluon correction 
$b\to c\bar{c}[n]+q+g$, where the $c\bar{c}$ pair is projected on one 
of the states in (\ref{set}). The corresponding diagrams are shown 
in Figs.~\ref{figa} and \ref{figb} respectively. The decay rate 
into a quarkonium can be written as the sum of partial decay rates 
through one of the intermediate $c\bar{c}$ states $n$. At 
next-to-leading order the partial decay rates take the form
\begin{eqnarray}
\label{notdecay}
\Gamma[n] &=& \Gamma_0\,\Big[C_{[1,8]}^2 f[n](\eta)\left(1+\delta_P[n]
\right) \nonumber\\
&&\hspace*{-1cm}
+ \,
\frac{\alpha_s(\mu)}{4\pi}\left(C_{[1]}^2 g_1[n](\eta) + 2 C_{[1]} C_{[8]} 
g_2[n](\eta) + C_{[8]}^2 g_3[n](\eta) \right)\Big]\,
\langle{\cal O}^H[n]\rangle,
\end{eqnarray}
where 
\begin{equation}
\label{gamma0}
\Gamma_0= \frac{G_F^2 |V_{bc}|^2 m_b^3}{216\pi (2 m_c)},
\end{equation}
and $\eta =4 m_c^2/m_b^2$. The operators ${\cal O}^H[n]$ are defined 
as in Ref.~\cite{BBL95}. The LO term is multiplied by 
$C_{[1]}^2$ if $n$ is a colour singlet state and by 
$C_{[8]}^2$ if $n$ is a colour octet state. We also 
used the fact that $|V_{cs}|^2+|V_{cd}|^2\approx 1$ 
to high accuracy. The functions $f[n]$ and $g_i[n]$ 
will be given later. The LO contribution is multiplied by a correction 
term $\delta_P[n]$ due to the penguin operators in (\ref{eq:Heff}). 
Likewise, we write the quarkonium energy 
distribution as 
\begin{eqnarray}
\label{notedist}
\frac{d\Gamma[n]}{dx}&=&\Gamma_0\,\Big[C_{[1,8]}^2 
f[n](\eta)\left(1+\delta_P[n]\right)\, \delta(1+\eta-x)
\nonumber\\
&&\hspace*{-1cm}
+\,\frac{\alpha_s(\mu)}{4\pi}\left(C_{[1]}^2 
g_1[n](\eta,x) + 2 C_{[1]} C_{[8]} 
g_2[n](\eta,x) + C_{[8]}^2 g_3[n](\eta,x) \right)\Big]\,
\langle{\cal O}^H[n]\rangle
\end{eqnarray}
where $x=2P\cdot p_b/m_b^2$. Note that to leading order in 
$\Lambda_{QCD}/m_b$ we do not distinguish the $b$ quark mass from 
the $B$ meson mass. To the order in the velocity expansion considered 
in this paper, we can also identify the momentum of the quarkonium 
with the momentum $P$ of the $c\bar{c}$ pair. (The kinematic effect 
of distinguishing the two is discussed in Ref.~\cite{BRW97}.) Hence 
$x$ can also be identified with $2 E_H/M_B$, where 
$E_H$ is the quarkonium energy in the $B$ meson rest frame and 
$M_B$ the $B$ meson mass. 

We now discuss which intermediate $c\bar{c}$ states should be taken 
into account for the production of a given quarkonium $H$.

\boldmath $J/\psi$, $\psi'$\unboldmath: 
At leading order in the velocity expansion the 
spin-triplet $S$-wave charmonium states are produced directly from 
a $c\bar{c}$ pair with the same quantum numbers, i.e. 
$n={}^3\!S_1^{(1)}$. At order $v^4$ relative to this colour singlet 
contribution, a $\psi$ can materialize through the colour octet 
$c\bar{c}$ states $n={}^3\!S_1^{(8)}, {}^1\!S_0^{(8)}, 
{}^3\!P_J^{(8)}$, where the subscript `$J$' implies a sum over 
$J=0,1,2$. The suppression factor $v^4$ follows from the counting 
rules for the multipole transitions for soft gluons that convert 
the state $n$ into the $\psi$ meson \cite{BBL95}. 
The leading order colour singlet 
contribution is proportional to $C_{[1]}^2$, while the colour 
octet terms are proportional to $C_{[8]}^2$. 
Because the weak effective 
Hamiltonian favours the production of colour octet $c\bar{c}$ pairs by 
a large factor
\begin{equation}
C_{[8]}^2/C_{[1]}^2\approx 15,
\end{equation}
the colour octet contributions must be included, since their 
suppression by $v^4\sim 1/15$ (for $J/\psi$) can easily be compensated. 
(The numbers serve only as order of magnitude estimates of the 
relative importance of the colour singlet and the colour octet 
contributions.) According to the velocity 
counting rules, there is a correction of order 
$v^2$ to the colour singlet contribution related to the derivative operator  
${\cal P}_{1}({}^3\!S^{(1)})$ as defined in Ref.~\cite{BBL95}. Because it is 
multiplied by the small coefficient $C_{[1]}^2$, and because we will find 
that the colour singlet contribution is indeed a small contribution 
to the total production cross section, we do not 
consider this additional correction in what follows. Similar derivative 
operators contribute to the colour octet channels. We do not take them 
into account, because we do not take into account other corrections of order 
$v^6$ with the large coefficient $C_{[8]}^2$. Hence, even after 
including the NLO correction in $\alpha_s$, there remains an uncertainty 
of order $v^2\sim 25\%$ in the theoretical prediction, assuming that the 
long-distance matrix elements were accurately known.

\boldmath $\eta_c$\unboldmath: 
The same discussion applies to the spin-singlet state. The 
colour singlet contribution involves $n={}^1\!S_0^{(1)}$. At  
relative order $v^4$, $\eta_c$ can be produced through the colour octet 
$c\bar{c}$ states $n={}^1\!S_0^{(8)}, {}^3\!S_1^{(8)}, 
{}^1\!P_1^{(8)}$.

\boldmath $\chi_{cJ}$\unboldmath: 
At leading order in the velocity expansion, both 
$n={}^3\!P_J^{(1)}$ and $n={}^3\!S_1^{(8)}$ contribute to the production 
of a the spin-triplet $P$-wave state \cite{BBYL92}. 
Because the partial production 
rate through the ${}^3\!S_1^{(8)}$ state is already multiplied by the 
large coefficient $C_{[8]}^2$, it is not necessary to go to higher 
orders in the velocity expansion. Note that, because of the $V-A$ structure 
of the weak vertex, a $c\bar{c}$ pair cannot be produced in a 
${}^3\!P_{0,2}$ angular momentum state at LO in $\alpha_s$.

\boldmath $h_c$\unboldmath: 
The same discussion as for the $\chi_{cJ}$ states applies 
to the spin-singlet $P$-wave state. In this case we take into account 
$n={}^1\!P_1^{(1)}$ and $n={}^1\!S_0^{(8)}$ at NLO in $\alpha_s$.
Owing to the $V-A$ structure 
of the weak vertex, a $c\bar{c}$ pair cannot be produced in a 
${}^1\!P_{1}$ angular momentum state at LO in $\alpha_s$. 

\section{Outline of the calculation}
\label{calculation}

\noindent
The Feynman diagrams shown in Figs.~\ref{figa} and \ref{figb} are 
projected onto a colour and angular momentum state as specified in 
(\ref{set}). The virtual corrections contain ultraviolet (UV) 
divergences, which can be absorbed into a renormalization of the 
operators ${\cal O}_{1,8}$ in the weak effective Hamiltonian 
(\ref{eq:Heff}). The virtual corrections 
contain infrared (IR) divergences, which cancel against IR divergences 
in the real corrections. In addition, the real corrections contain 
IR divergences due to the emission of soft gluons from the $c$ or 
$\bar{c}$ lines, which do not cancel with IR divergences in the virtual 
correction, if the $c\bar{c}$ pair is projected on a $P$-wave 
state. These IR divergences can be factorized and absorbed into a 
renormalization of the non-perturbative matrix elements 
$\langle {\cal O}^H[n]\rangle$. In the following we provide some 
details on the UV and IR regularization, which are specific of the 
present calculation. More details on the strategy of a next-to-leading 
order calculation can be found in Ref.~\cite{PCGMM98}, which deals 
with quarkonium decay and total quarkonium production cross 
sections in fixed-target collisions.

\subsection{UV regularization and the treatment of $\gamma_5$}
\label{uvreg}

\noindent
The UV divergences are regulated dimensionally and the IR divergences 
are regulated with a gluon mass. 
The UV divergences in the diagrams of Fig.~\ref{figa} 
cancel against the UV divergences in diagrams (not shown in the figure) 
with the insertion of 
the 1-loop counterterm for ${\cal O}_{1,8}$. 
We combine the diagram with its counterterm 
diagram before projecting on a particular $c\bar{c}$ state $n$, and 
before taking the 2-particle phase space integral. This has the advantages 
that it avoids extending the projection to $d$ dimensions and that 
the phase space integral can also be done in four dimensions. 

The finite part of the virtual gluon correction depends on the 
prescription for handling $\gamma_5$ in $d$ dimensions. This has to be 
chosen consistently with the one used to define the operators 
${\cal O}_{1,8}$ in Ref.~\cite{BW90}. The prescription consists of a 
definition of $\gamma_5$ and its anti-commutation property, 
together with a choice of `evanescent operators'. The evanescent 
operators are 
implicitly defined by specifying the order $\epsilon$ 
(where $d=4-2\epsilon$) terms of the following products of 
Dirac matrices:
\begin{eqnarray}
\gamma_\rho\gamma_\alpha\Gamma_\mu \otimes 
\gamma_\rho\gamma_\alpha\Gamma_\mu 
&=& (16+4 X_R\epsilon)\,\Gamma_\mu \otimes 
\Gamma_\mu + E_X
\\
\Gamma_\mu\gamma_\rho\gamma_\alpha \otimes 
\gamma_\alpha\gamma_\rho\Gamma_\mu 
&=& (4+4 Y_R\epsilon)\,\Gamma_\mu \otimes 
\Gamma_\mu + E_Y 
\\
\Gamma_\mu\otimes
\gamma_\rho\gamma_\alpha\Gamma_\mu\gamma_\alpha\gamma_\rho 
&=& (4+4 Z_R\epsilon)\,\Gamma_\mu \otimes 
\Gamma_\mu + E_Z.
\end{eqnarray}
(Here we defined $\Gamma_\mu=\gamma_\mu(1-\gamma_5)$.) The renormalization 
conventions of Refs.~\cite{BW90,CFMR94} correspond to:
\begin{eqnarray}
\mbox{NDR scheme}: && \quad X_{\rm NDR}=-1\qquad Y_{\rm NDR}=Z_{\rm NDR}=-2,\\
\mbox{HV scheme\hspace*{0.1cm}}: && \,\,\quad X_{\rm HV}=-1\qquad\,\,\,\,\, 
Y_{\rm HV}=Z_{\rm HV}=0.
\end{eqnarray}
In the HV scheme, vertex diagrams are treated differently in 
Ref.~\cite{BW90} and Ref.~\cite{CFMR94}. As a consequence, as already 
mentioned above, in the HV scheme one has to multiply the 
coefficients $C_{\pm}(\mu)$ defined in refs.~\cite{BBL96,BW90} by the 
factor $1-4\alpha_s(\mu)/(3\pi)$, while this factor is already included 
in the definition of Ref.~\cite{CFMR94}. We checked that our final 
result is identical in the NDR and HV schemes up to terms beyond 
NLO accuracy, if we use the expressions 
for $C_{\pm}(\mu)$ of Sect.~\ref{notation} including the 
additional factor just mentioned in the HV scheme.

The coefficient functions quoted in Sect.~\ref{notation} refer to 
a Fierz version of the weak Hamiltonian different from 
(\ref{eq:Heff}) and Fierz transformations do not commute with 
renormalization in general. If we use the standard Fierz version 
rather than the singlet-octet form quoted in (\ref{eq:Heff}), this 
interchanges $Y_R$ and $Z_R$ in the results quoted in Appendix~A.3. 
However, since in both schemes we used one has $Y_R=Z_R$, either of the 
two Fierz versions can be used.

The NLO calculation for $n={}^3\!S_1^{(1)}$ has already been done 
in Ref.~\cite{BE94} in the HV scheme. We find that our 
result for the functions $g_i[{}^3\!S_1^{(1)}]$ defined in 
(\ref{notdecay}) and given in the Appendix agrees with 
the result of Ref.~\cite{BE94}. Nevertheless our result for the 
contribution of this channel to the decay rate, given by 
$C_{[1]}^2\,f[{}^3\!S_1^{(1)}](\eta)\,+\,$NLO terms, differs from 
the one given in Ref.~\cite{BE94}, because the authors 
of Ref.~\cite{BE94} used the coefficient functions of Ref.~\cite{BW90},
but did not correct them (or alternatively, the low energy 
matrix elements) 
by the factor $1-4\alpha_s(\mu)/(3\pi)$. As explained above 
with the conventions of Ref.~\cite{BW90} this additional 
factor is necessary in the HV scheme to obtain a scheme-independent 
result.

\subsection{IR regularization and NRQCD factorization}

\noindent
The real and virtual corrections individually have double-logarithmic 
IR divergences, which we regulate by a gluon mass. However, after 
adding all contributions to the partonic process $b\to c\bar{c}+X$, 
the IR divergences do not cancel completely. The remaining 
IR divergences are associated only with emission from the $c$ and 
$\bar{c}$ quark. This is a necessary (but not sufficient) requirement 
for their factorization into NRQCD matrix elements as discussed 
in detail in Ref.~\cite{BBL95}. 

In addition to these IR divergences related to soft gluon emission, 
the last diagram in Fig.~\ref{figa} exhibits the well-known 
Coulomb divergence, when the relative momentum of the $c$ and 
$\bar{c}$ is set to zero. We regularize this divergence by keeping 
the relative momentum finite in the integrals, which would otherwise give 
rise to the Coulomb singularity.

In order to extract the short-distance parts $\Gamma[m]$ (see 
(\ref{notdecay})) of the 
partonic decay, we write 
\begin{equation}
\label{par}
\Gamma(b\to c\bar{c}[n]+X) = \sum_m
\Gamma[m]\,\langle{\cal O}^{\rm pert}[m]
\rangle,
\end{equation}
where $\langle{\cal O}^{\rm pert}[m]\rangle$ denotes the NRQCD matrix element 
for a perturbative $c\bar{c}$ pair in the state $n$. 
At NLO one has to calculate the left-hand side 
and $\langle{\cal O}^{\rm pert}[m]\rangle$ to NLO.

\begin{figure}[t]
   \vspace{-4cm}
   \epsfysize=23cm
   \epsfxsize=15.6cm
   \centerline{\epsffile{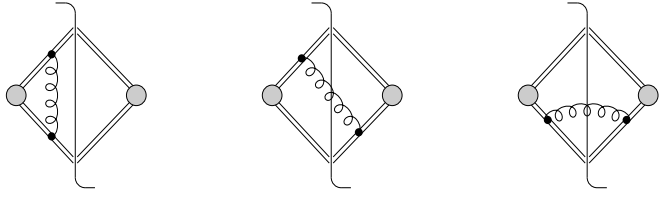}}
   \vspace*{-15.3cm}
\caption[dummy]{Perturbative corrections to the NRQCD operator 
matrix elements. A shaded circle denotes an insertion that 
specifies the angular momentum and colour of the $c\bar{c}$ pair. 
The vertical line implies that the diagram is `cut'. Symmetric 
diagrams are not shown.
\label{figc}}
\end{figure}

The diagrams that contribute the $\alpha_s$ correction to 
$\langle{\cal O}^{\rm pert}[m]\rangle$ are shown in Fig.~\ref{figc}. 
For the first diagram (together with its complex conjugate) we 
obtain
\begin{equation}
\langle {\cal O}^{\rm pert,1}[n]\rangle_{\rm A} = 
\langle {\cal O}^{\rm pert,0}[n]\rangle \cdot A[n]\,
 a_s\cdot\frac{2\pi^2}{v},
\end{equation}
where $A[n]=C_F=4/3$, if $n$ is a colour singlet state,  
$A[n]=-1/(2 N_c)=-1/6$ if $n$ is a colour octet state and $v$ is the 
relative velocity of the two quarks. (The superscript `0' refers to  
a matrix element at tree level, `1' denotes a 1-loop 
contribution.) This renders the short-distance 
coefficients free of the Coulomb singularity.

%%%%%%%%%%%%%%%%%%%%%%%%%%%%%%%%%%%%%%%%%%%%%%%%%%%%%%%%%%%%%%%%%%%
%%%%%%%%%%%%%%%%%%%%%%%%%%%%%%%%%%%%%%%%%%%%%%%%%%%%%%%%%%%%%%%%%%%
The other two diagrams (called collectively `B') together with their 
symmetry partners are UV and IR divergent. We define the NRQCD matrix 
elements in the $\overline{\rm MS}$ scheme and denote 
their renormalization scale by $\hat{\mu}$. The IR divergence is regulated 
with a gluon mass to be consistent with the IR regulator used for the 
evaluation of the partonic process on the left-hand side of (\ref{par}). 
The result is (compare with the Appendix of Ref.~\cite{BBL95} and with 
Ref.~\cite{PCGMM98}, where other IR regulators are used):
\begin{eqnarray}
\label{rgeir}
\langle {\cal O}^{\rm pert,1}_1({}^3 \!S_1)\rangle_{\rm B} &=&
\frac{\alpha_s}{4\pi}\left(\ln\frac{\lambda^2}{\hat{\mu}^2}+\frac{1}{3}\right) 
\left(-48\right) \langle {\cal O}^{\rm pert,0}_8({}^3 \!P_0)\rangle/m_c^2,
\nonumber\\
\langle {\cal O}^{\rm pert,1}_8({}^3 \!S_1)\rangle_{\rm B} &=&
\frac{\alpha_s}{4\pi}\left(\ln\frac{\lambda^2}{\hat{\mu}^2}+\frac{1}{3}\right) 
\bigg[\left(-\frac{32}{3}\right) \langle {\cal O}^{\rm pert,0}_1({}^3 \!P_0)
\rangle/m_c^2
\nonumber\\
&&\hspace*{-1.5cm}+\,
\left(-20\right) \langle {\cal O}^{\rm pert,0}_8({}^3 \!P_0)
\rangle/m_c^2\bigg],
\nonumber\\
\langle {\cal O}^{\rm pert,1}_1({}^1 \!S_0)\rangle_{\rm B} &=&
\frac{\alpha_s}{4\pi}\left(\ln\frac{\lambda^2}{\hat{\mu}^2}+\frac{1}{3}\right) 
\left(-\frac{16}{3}\right) \langle {\cal O}^{\rm pert,0}_8({}^1 \!P_1)
\rangle/m_c^2,
\\
\langle {\cal O}^{\rm pert,1}_8({}^1 \!S_0)\rangle_{\rm B} &=&
\frac{\alpha_s}{4\pi}\left(\ln\frac{\lambda^2}{\hat{\mu}^2}+\frac{1}{3}\right) 
\bigg[\left(-\frac{32}{27}\right) \langle {\cal O}^{\rm pert,0}_1({}^1 \!P_1)
\rangle/m_c^2\nonumber\\
&&\hspace*{-1.5cm}
+\,\left(-\frac{20}{9}\right) \langle {\cal O}^{\rm pert,0}_8({}^1 \!P_1)
\rangle/m_c^2\bigg].
\nonumber
\end{eqnarray}
($\lambda$ denotes the gluon mass.)
Note that if one breaks 
up the $^3\!P$ term into terms with different $J$, one should replace 
\begin{equation}
\langle {\cal O}^{\rm pert,0}_{1,8}({}^3 \!P_0)\rangle \to 
1/9\,(\langle {\cal O}^{\rm pert,0}_{1,8}({}^3 \!P_0)\rangle + 
\langle {\cal O}^{\rm pert,0}_{1,8}({}^3 \!P_1)\rangle + 
\langle {\cal O}^{\rm pert,0}_{1,8}({}^3 \!P_2)\rangle).
\end{equation}
Using these results and solving for $\Gamma[n]$ we find the IR finite 
short-distance coefficients for each $n$ collected in Appendix~A.3.

\subsection{Difficulties with the colour singlet channels}
\label{diffsing}

\noindent
The LO contributions to the colour singlet channels ${}^1\!S_0^{(1)}$, 
${}^3\!S_1^{(1)}$ and ${}^3\!P_1^{(1)}$ are proportional to the 
small and strongly scale dependent coefficient $C_{[1]}^2(\mu)$. 
One would therefore expect the NLO contribution to be particularly 
important for these channels. However, the strict NLO calculation leads 
to a negative, and therefore meaningless decay rate into these channels 
and to the conclusion that a reliable result can only be obtained 
at next-to-next-to-leading order. This problem was already identified 
and discussed in Ref.~\cite{BE94}. (For the remainder of this section 
it is assumed that the reader has consulted Ref.~\cite{BE94} for more 
details.) 

Consider the three next-to-leading order terms $g_i$ in 
(\ref{notdecay}). Despite its large coefficient $C_{[8]}^2$ the 
$g_3$-term, which comes only from a real correction, turns out to be 
numerically very small (see the tables in the following section). 
Both $g_1$ and $g_2$ (at $\mu=m_b$) are large 
and negative, and $g_2$ in particular, which comes with the larger 
coefficient $2 C_{[1]} C_{[8]}$, drives the decay rate negative.

The authors of Ref.~\cite{BE94} suggested treating the decay process 
in a simultaneous expansion in $\alpha_s$ and $C_{[1]}/C_{[8]}$. This 
implies that one should add to the term of order 
$\alpha_s C_{[1]} C_{[8]}$ all terms of order 
$\alpha_s^2 C_{[8]}^2$, because they also count as NLO in this 
rearranged expansion. On the other 
hand, the term $\alpha_s C_{[1]}^2$ (which involves $g_1$) should be 
neglected as being of higher order. The authors of Ref.~\cite{BE94} 
did not actually calculate all terms of order $\alpha_s^2 C_{[8]}^2$, 
but estimated them by adding 
\begin{equation}
\label{impro}
\Gamma_0\,\langle{\cal O}^H[n]\rangle \left(\frac{\alpha_s(\mu)}{4\pi}
\right)^2\,C_{[8]}^2\,\frac{g_2[n]^2}{f[n]}
\end{equation}
to (\ref{notdecay}). This estimate can be motivated as follows: the 
virtual contribution to $g_2$ is given by the first four diagrams 
in Fig.~\ref{figa} times the (complex conjugated) tree amplitude. 
All two-particle $[c\bar{c}]q$ cuts to the $\alpha_s^2 C_{[8]}^2$ 
term are given by the square of the first four diagrams 
in Fig.~\ref{figa}. Hence, ignoring the real contribution, one may 
argue that $g_2[n]^2/f[n]$ is close (but not equal) to the two-particle 
contributions to the $\alpha_s^2 C_{[8]}^2$ term. 

In Ref.~\cite{ST97} the square of the 1-loop amplitude with a 
$[c\bar{c}]q$ final state is computed exactly and argued to provide 
a better estimate than the original one of Ref.~\cite{BE94}, because 
one leaves out only real contributions to the coefficient of 
$\alpha_s^2 C_{[8]}^2$, which are argued to be phase-space 
suppressed. However, we find that for the ${}^3\!P_1^{(1)}$ channel 
the virtual contributions alone are IR divergent. Therefore the 
real correction that cancels this divergence cannot be argued to 
be small. In our opinion this also calls into question the assumption 
that the real contributions are numerically small for the $S$-wave 
channels. For this reason we choose to follow 
with a minor modification the procedure 
of Ref.~\cite{BE94}, which adds an IR finite term by construction, 
since $g_2$ is IR finite. The minor modification is the following: 
the third and fourth diagrams in Fig.~\ref{figa} have imaginary 
parts, which contribute to the real part of the square of the 
amplitude (and hence to the coefficient of $\alpha_s^2 C_{[8]}^2$). 
The remnants of these imaginary parts after multiplying 
the one-loop amplitude by the complex conjugate of the tree amplitude 
can easily be restored from the  
$\ln(1-\eta)$-term (and $\mbox{Li}_2(\eta)$-term in the case of 
${}^3\!P_1^{(1)}$) in the results presented in Appendix~A.3. If we call 
the restored imaginary part $\mbox{Im}\,g_2$, then we use 
(\ref{impro}), with $g_2^2$ replaced by $g_2^2+(\mbox{Im}\,g_2)^2$.

We wish to emphasize two points: first, the discussed modifications 
of the colour singlet channels are certainly ad hoc and 
should be regarded with great caution. Second, the effect on the 
decay rate into a particular quarkonium state is not severely 
affected by this uncertainty, because it is dominated by colour 
octet contributions, whose short-distance coefficients can be 
computed reliably at NLO as we shall see. 

In order to gain a numerical understanding 
of the importance of the various terms involved in the colour 
singlet channels, we consider in the following three 
computational schemes for the decay rate: (a) the (strict) NLO calculation; 
(b) the NLO calculation with the term (\ref{impro}) added, but 
without the $g_1$-term (`improved'); (c) the same as (b), but with 
$g_1$ included (`total'). For the $S$-wave colour singlet channels, (a) and 
(c) yield a negative rate. They are therefore meaningless. Option (b) 
yields a positive result of a magnitude similar to the result 
of Refs.~\cite{BE94,ST97}. It may be considered as an order-of-magnitude 
estimate for the colour singlet contribution, but it may well be 
uncertain by 100\%. For the ${}^3\!P_1^{(1)}$ channel all three 
options give negative partial rates. However, since this 
channel mixes with ${}^3\!S_1^{(8)}$ and only the sum of the two is physical, 
a negative partial rate is not unphysical by itself.

\section{Results and Discussion}
\label{results}

\noindent
In this section we present our results for the branching fractions 
of $B$ decay into charmonium and moments of the quarkonium energy 
distributions in numerical form. The analytic expressions that enter 
(\ref{notdecay}) and (\ref{notedist}) are collected in the appendices 
for reference. 

\subsection{Branching ratios for $B$ decay into charmonium}

\subsubsection{General discussion of NLO corrections}

\noindent
We normalize our calculation to the theoretical expression for the 
inclusive semileptonic decay rate
\begin{equation}
\label{semi}
\Gamma^{\rm th}_{\rm SL}= \frac{G_F^2 |V_{bc}|^2 m_b^5}{192 \pi^3}
\left(1 - 8 z^2 + 8 z^6 - z^8 - 24 z^4 \ln z\right)
\eta_1(z),
\end{equation}
where $z=m_c/m_b$ and 
\begin{equation}
\eta_1(z) = 1-\frac{2\alpha_s(m_b)}{3\pi}
\left(\frac{3}{2} + \left(-\frac{31}{4} + \pi^2\right)\left(1 - z\right)^2 
\right)
\end{equation}
represents an excellent approximation \cite{corbo} for the 
1-loop QCD correction factor. (The complete analytic result can be found 
in Ref.~\cite{nir}.) 
For any particular quarkonium state $H$, we obtain the 
branching fraction in the form\footnote{When $n$ is a $P$-wave state, 
$\langle{\cal O}^H[n]\rangle$ should be understood as 
$\langle{\cal O}^H[n]\rangle/m_c^2$ in the following formula, so  
that all matrix elements have mass dimension 3. Furthermore, 
in the case of $n={}^3\!P_J^{(8)}$ which refers to the $c\bar{c}$ state 
${}^3\!P^{(8)}$ summed over $J=0,1,2$, the NRQCD matrix element 
$\langle {\cal O}^H[n]\rangle$ is chosen to be 
$\langle {\cal O}^H_8({}^3\!P_0)\rangle/m_c^2$.}
\begin{eqnarray}
\label{br}
\mbox{Br}\,(B\to H+X) &=& {\cal N}\,
\sum_n \,\langle{\cal O}^H[n]\rangle
\Big[C_{[1,8]}^2 f[n](\eta)\left(1+\delta_P[n]
\right) \nonumber\\
&&\hspace*{-1cm}
+ \,
\frac{\alpha_s(\mu)}{4\pi}\left(C_{[1]}^2 g_1[n](\eta) + 2 C_{[1]} C_{[8]} 
g_2[n](\eta) + C_{[8]}^2 g_3[n](\eta) \right)\Big].
\end{eqnarray}
The overall factor is given by
\begin{equation}
{\cal N} = \mbox{Br}^{\rm exp}_{\rm SL}\,\frac{\Gamma_0}
{\Gamma^{\rm th}_{\rm SL}}=3.0\cdot 10^{-2}\,{\rm GeV}^{-3},
\end{equation}
where we used $\mbox{Br}^{\rm exp}_{\rm SL}=10.4\%$ and $\Gamma_0$ as 
given by (\ref{gamma0}). The charm 
and bottom pole masses are taken to be 
$1.5\,$GeV and $4.8\,$GeV, respectively. This yields 
$\eta=0.39$, which we use unless otherwise mentioned. The sensitivity 
of the charmonium production cross sections to 
the quark mass values will be discussed below.

We first examine the impact of the next-to-leading order correction 
and the dependence on the factorization scale $\mu$ for each 
intermediate $c\bar{c}$ state separately. We neglect the penguin 
contribution for this purpose. In Table~\ref{tabcomp} we show the 
branching fractions excluding the dimensionless normalization factor 
${\cal N} \langle {\cal O}^H[n] \rangle$ for three values of $\mu$ 
at LO and at NLO. To evaluate the LO expression we also use the 
Wilson coefficients at LO and 1-loop running of the strong coupling 
with $\Lambda_{QCD}^{LO}$ such that $\alpha_s(M_Z)=0.119$ both in 
LO and NLO. This is a large effect for the colour singlet channel, 
since $C_{[1]}(m_b)=0.55$, $C_{[8]}(m_b)=2.14$ (in the NDR scheme) but 
$C_{[1]}^{LO}(m_b)=0.41$, $C_{[8]}^{LO}(m_b)=2.19$.

\begin{table}[t]
\addtolength{\arraycolsep}{0.25cm}
\renewcommand{\arraystretch}{1.4}
$$
\begin{array}{c|c|c|c|c|c|c}
\hline\hline
n & \multicolumn{3}{|c|}{\mbox{LO}} & \multicolumn{3}{|c}{\mbox{NLO}}\\ 
\hline 
\mu/\mbox{GeV} & 2.5 & 5 & 10 & 2.5 & 5 & 10 \\ 
\hline
^1\!S_0^{(1)} & 0.0453 & 0.201 & 0.407 & -0.219 & -0.426 & -0.554 \\
^3\!S_1^{(1)} & 0.0269 & 0.119 & 0.242 & -0.119 & -0.250 & -0.334 \\
^3\!P_0^{(1)} & 0      & 0     & 0     & -0.660 & -0.481 & -0.377 \\
^3\!P_1^{(1)} & 0.0537 & 0.238 & 0.484 & -0.654 & -0.738 & -0.794 \\
^3\!P_2^{(1)} & 0      & 0     & 0     & -0.534 & -0.389 & -0.305 \\
^1\!P_1^{(1)} & 0      & 0     & 0     & -1.02  & -0.741 & -0.58  \\ 
\hline
^1\!S_0^{(8)} & 8.72 & 8.01 & 7.51 & 12.6 & 11.1 & 10.2 \\
^3\!S_1^{(8)} & 5.18 & 4.75 & 4.46 & 7.70 & 6.80 & 6.18 \\
^3\!P_J^{(8)} & 31.1 & 28.5 & 26.8 & 38.3 & 34.5 & 31.7 \\
^1\!P_1^{(8)} & 0    & 0    & 0    & -1.95 & -1.53 & -1.30 \\
\hline\hline
\end{array}
\vspace*{0.2cm}
$$
\caption[dummy]{\label{tabcomp}
Comparison of LO and NLO for the decay 
rate into a $c\bar{c}$ pair in state $n$ and the dependence on the 
factorization scale $\mu$. The dimensionless overall factor 
${\cal N}\, \langle {\cal O}^H[n] \rangle$ is not included. Quark masses:
$m_b=4.8\,$GeV, $m_c=1.5\,$GeV. Penguin contribution not included.}
\end{table}
\begin{table}[t]
\addtolength{\arraycolsep}{0.25cm}
\renewcommand{\arraystretch}{1.4}
$$
\begin{array}{c|c|c|c|c}
\hline\hline
n & \mu & ^1\!S_0^{(1)} & ^3\!S_1^{(1)} & ^3\!P_1^{(1)} \\ 
\hline 
           & 2.5 & 0.0453 & 0.0269 & 0.054  \\
\mbox{LO}  & 5   & 0.201  & 0.119  & 0.238  \\
           & 10  & 0.407  & 0.242  & 0.484   \\
\hline
           & 2.5 & -0.219 & -0.119 & -0.654  \\
\mbox{NLO}  & 5  & -0.426 & -0.250 & -0.738  \\
           & 10  & -0.554 & -0.334 & -0.794  \\
\hline
           & 2.5 & 0.0951 & 0.034 & -0.392  \\
\mbox{`Impr'}  & 5   & 0.058  & 0.0259  & -0.280  \\
           & 10  & 0.0706  & 0.0402  & -0.176   \\
\hline
           & 2.5 & 0.0283 & -0.0138 & -0.455  \\
\mbox{`Tot'}  & 5   & -0.0984  & -0.0862  & -0.427  \\
           & 10  & -0.159  & -0.124  & -0.391   \\
\hline\hline
\end{array}
\vspace*{0.2cm}
$$
\caption[dummy]{\label{tabsing} The same quantity as 
in Table~I for various treatments of the colour 
singlet channel (for which the LO short distance coefficient, i.e. 
$f[n]$, is non-vanishing). Penguin contribution not included.}
\end{table}

We now observe that the colour singlet contributions 
are, as expected, enormously scale-dependent at LO. The NRQCD matrix element 
$\langle {\cal O}_1^{J/\psi}({}^3\!S_1)\rangle$ is related to the 
radial wavefunction at the origin by 
$\langle {\cal O}_1^{J/\psi}({}^3\!S_1)\rangle=9|R(0)|^2/(2\pi)$ 
up to corrections of order $v^4$. Using 
$\langle {\cal O}_1^{J/\psi}({}^3\!S_1)\rangle=1.16\,$GeV$^3$ 
\cite{EQ95}, we obtain
\begin{equation}
\label{losing}
\mbox{Br}\,(B\to J/\psi+X) = (0.09 - 0.84)\%\qquad\mbox{(colour singlet, 
LO)}
\end{equation}
to be compared with the measured branching 
fraction $(0.80\pm0.08)\%$ \cite{CLEO95}.\footnote{Note 
that we denote by $\mbox{Br}\,(B\to J/\psi+X)$ the direct 
production of $J/\psi$, excluding radiative decays into $J/\psi$ 
from higher-mass charmonium states. The same convention applies to 
all other charmonium states $H$.} 
The LO prediction is uncertain by a factor of about 10 for 
all colour singlet channels, as can be seen from Table~\ref{tabcomp}. 
As also seen from this table, the scale uncertainty is reduced to a 
factor 2--3 at NLO. However, the NLO correction term renders the 
partial decay rates negative, as already mentioned in 
Section~\ref{diffsing}.

The situation can be somewhat improved by adding the estimate 
(\ref{impro}) for the 
order $\alpha_s^2$ NNLO term with the large coefficient $C_{[8]}^2$, 
while treating the $\alpha_s C_{[1]}^2$ term as formally of higher 
order in a double expansion in $\alpha_s$ and $C_{[1]}/C_{[8]}$ 
\cite{BE94}. The addition of (\ref{impro}) also reduces the 
factorization scale dependence further, because it contains exactly 
the double logarithmic correction 
$\alpha_s^2 C_{[8]}^2 \ln^2(m_b^2/\mu^2)$, which is required to 
cancel the large scale dependence of $C_{[1]}^2$ at leading order 
in $\alpha_s$. In Table~\ref{tabsing} we display the result for the 
partial decay rates into the colour singlet channel, which is obtained in 
this way (denoted `Impr' in the table) and for comparison again the 
LO and NLO result. The improvement can and should be done only 
for those colour singlet channels that have non-vanishing LO 
contributions. The last three rows of Table~\ref{tabsing} show the results  
that are obtained if we add back the $g_1$ term in (\ref{br}) to 
the improved treatment. The $\alpha_s C_{[1]}^2 g_1$ term is 
sizeable and negative and therefore re-introduces a large scale 
dependence. The same improvement that is applied 
to the LO $C_{[1]}^2 f$ term is necessary for the $g_1$ term, 
which would require going to order $\alpha_s^3 C_{[8]}^2$. One 
may argue that unless this is done, it is preferable to leave 
the $g_1$ term out entirely. Therefore we shall use the 
`improved' version (`Impr' in Table~\ref{tabsing}) as our 
default option later. While the result is certainly not accurate, 
we believe that this is the best we can do to the colour singlet 
channel without making arbitrary modifications. We note that for 
$J/\psi$ this gives a colour singlet contribution to the branching 
fraction, which is close to the lower limit in (\ref{losing}) and also 
compatible with the estimates of Refs.~\cite{BE94,ST97}. It 
seems safe to conclude that colour singlet 
production alone is not sufficient to explain 
the measured branching fraction.

The partial rates in the four relevant colour octet channels are 
shown in the lower part of Table~\ref{tabcomp}. In this case we 
find that the perturbative expansion is very well behaved. The NLO 
short-distance coefficients are larger by 20\%--50\% than the LO 
coefficients and the scale dependence is very moderate. The scale 
dependence is not reduced from LO to NLO. This is due to the fact 
that the LO coefficients depend only on the scale-insensitive 
$C_{[8]}$, while there are sizeable coefficients of the highly 
scale-dependent combinations $C_{[1]} C_{[8]}$ and $C_{[1]}^2$ 
at NLO. The numerical enhancement of the short-distance coefficients 
in the colour octet channels, which is evident from Table~\ref{tabcomp}, is 
sufficient to account for 
the measured $J/\psi$ branching fraction, as already noted 
in Refs.~\cite{KLS96,FHMN97}. The positive NLO correction 
reinforces this trend. Other $J/\psi$ production processes suggest 
that the long-distance parameters in the colour octet channels are 
of the order a few times $10^{-2}\,$GeV$^3$. (This will be made 
more precise soon.) This leads to typical branching fractions of 
order $0.5\%$.

\begin{table}[p]
\addtolength{\arraycolsep}{0.25cm}
\renewcommand{\arraystretch}{1.6}
$$
\begin{array}{c|c|c|c|c|c}
\hline\hline
n & f & \delta_P & g_1 & g_2 & g_3\\ 
\hline
^3\!S_1^{(1)} & 0.661 & -0.004  & -20.5 & -8.46+2.65\ln(m_b^2/\mu^2) & 
0.162 \\
^3\!S_1^{(8)} & 0.992 & -0.09   & 0.486 & -6.35+1.98\ln(m_b^2/\mu^2) & 
32.0-3.97\ln(m_b^2/\mu^2) \\
^1\!S_0^{(8)} & 1.67  & 0.009 & 0.556 & -11.2+3.34\ln(m_b^2/\mu^2) & 
50.2-6.68\ln(m_b^2/\mu^2) \\
^3\!P_J^{(8)} & 5.95  & 0.009 & -129  & -36.6+11.9\ln(m_b^2/\mu^2) & 
121-23.8\ln(m_b^2/\mu^2) \\
\hline\hline
\end{array}
$$
\caption[dummy]{\label{tab1} Numerical values for the LO and NLO 
functions (NDR scheme) that enter the branching 
ratio $\mbox{Br}\,(B\to J/\psi+X)$ according to (\ref{br}) for 
$m_b=4.8\,$GeV, $m_c=1.5\,$GeV. The estimate for the penguin correction 
$\delta_P$ is obtained with the parameters detailed in Appendix~A.2. 
The table applies without modification to $\psi'$.}
\end{table}

\begin{table}[p]
\addtolength{\arraycolsep}{0.25cm}
\renewcommand{\arraystretch}{1.6}
$$
\begin{array}{c|c|c|c|c|c}
\hline\hline
n & f & \delta_P & g_1 & g_2 & g_3\\ 
\hline
^1S_0^{(1)} & 1.11 & 0.07  & -28.6 & -15.0+4.46\ln(m_b^2/\mu^2) & 
0.185 \\
^1S_0^{(8)} & 1.67 & 0.009 & 0.556 & -11.2+3.34\ln(m_b^2/\mu^2) & 
50.2-6.68\ln(m_b^2/\mu^2) \\
^3S_1^{(8)} & 0.992  & -0.09 & 0.486 & -6.35+1.98\ln(m_b^2/\mu^2) & 
32.0-3.97\ln(m_b^2/\mu^2) \\
^1P_1^{(8)} & 0  & 0 & -28.2  & 0 & 
-17.4 \\
\hline\hline
\end{array}
$$
\caption[dummy]{\label{tab2} Numerical values for the LO and NLO 
functions (NDR scheme) that enter the branching 
ratio $\mbox{Br}\,(B\to \eta_c+X)$ according to (\ref{br}) for 
$m_b=4.8\,$GeV, $m_c=1.5\,$GeV. The estimate for the penguin correction 
$\delta_P$ is obtained with the parameters detailed in Appendix~A.2.}
\end{table}

\begin{table}[p]
\addtolength{\arraycolsep}{0.25cm}
\renewcommand{\arraystretch}{1.6}
$$
\begin{array}{c|c|c|c|c|c}
\hline\hline
n & f & \delta_P & g_1 & g_2 & g_3\\ 
\hline
^3P_0^{(1)} & 0 & 0  & 0 & 0 & 
-6.11 \\
^3S_1^{(8)} & 0.992 & -0.09 & 0.486 & -6.35+1.98\ln(m_b^2/\mu^2) & 
32.0-3.97\ln(m_b^2/\mu^2) \\
\hline\hline
\end{array}
$$
\caption[dummy]{\label{tab3} Numerical values for the LO and NLO 
functions (NDR scheme) that enter the branching 
ratio $\mbox{Br}\,(B\to \chi_{c0}+X)$ according to (\ref{br}) for 
$m_b=4.8\,$GeV, $m_c=1.5\,$GeV. The estimate for the penguin correction 
$\delta_P$ is obtained with the parameters detailed in Appendix~A.2.}
\end{table}

\begin{table}[p]
\addtolength{\arraycolsep}{0.25cm}
\renewcommand{\arraystretch}{1.6}
$$
\begin{array}{c|c|c|c|c|c}
\hline\hline
n & f & \delta_P & g_1 & g_2 & g_3\\ 
\hline
^3P_1^{(1)} & 1.32 & 0.07  & -26.8 & -16.3+5.29\ln(m_b^2/\mu^2) & 
-4.04 \\
^3S_1^{(8)} & 0.992 & -0.09 & 0.486 & -6.35+1.98\ln(m_b^2/\mu^2) & 
32.0-3.97\ln(m_b^2/\mu^2) \\
\hline\hline
\end{array}
$$
\caption[dummy]{\label{tab4} Numerical values for the LO and NLO 
functions (NDR scheme) that enter the branching 
ratio $\mbox{Br}\,(B\to \chi_{c1}+X)$ according to (\ref{br}) for 
$m_b=4.8\,$GeV, $m_c=1.5\,$GeV. The estimate for the penguin correction 
$\delta_P$ is obtained with the parameters detailed in Appendix~A.2.}
\end{table}

\begin{table}[p]
\addtolength{\arraycolsep}{0.25cm}
\renewcommand{\arraystretch}{1.6}
$$
\begin{array}{c|c|c|c|c|c}
\hline\hline
n & f & \delta_P & g_1 & g_2 & g_3\\ 
\hline
^3P_2^{(1)} & 0 & 0  & 0 & 0 & 
-4.94 \\
^3S_1^{(8)} & 0.992 & -0.09 & 0.486 & -6.35+1.98\ln(m_b^2/\mu^2) & 
32.0-3.97\ln(m_b^2/\mu^2) \\
\hline\hline
\end{array}
$$
\caption[dummy]{\label{tab5} Numerical values for the LO and NLO 
functions (NDR scheme) that enter the branching 
ratio $\mbox{Br}\,(B\to \chi_{c2}+X)$ according to (\ref{br}) for 
$m_b=4.8\,$GeV, $m_c=1.5\,$GeV. The estimate for the penguin correction 
$\delta_P$ is obtained with the parameters detailed in Appendix~A.2.}
\end{table}

\begin{table}[p]
\addtolength{\arraycolsep}{0.25cm}
\renewcommand{\arraystretch}{1.6}
$$
\begin{array}{c|c|c|c|c|c}
\hline\hline
n & f & \delta_P & g_1 & g_2 & g_3\\ 
\hline
^1P_1^{(1)} & 0 & 0  & 0 & 0 & 
-9.41 \\
^1S_0^{(8)} & 1.67 & 0.009 & 0.556 & -11.2+3.34\ln(m_b^2/\mu^2) & 
50.2-6.68\ln(m_b^2/\mu^2) \\
\hline\hline
\end{array}
$$
\caption[dummy]{\label{tab6} Numerical values for the LO and NLO 
functions (NDR scheme) that enter the branching 
ratio $\mbox{Br}\,(B\to h_c+X)$ according to (\ref{br}) for 
$m_b=4.8\,$GeV, $m_c=1.5\,$GeV. The estimate for the penguin correction 
$\delta_P$ is obtained with the parameters detailed in Appendix~A.2.}
\end{table}

For our standard value of the quark masses ($m_b=4.8\,$GeV, 
$m_c=1.5\,$GeV), the numerical values for the 
functions that enter (\ref{br}) are given in Tables~\ref{tab1}--\ref{tab6} 
for all of the six (known) charmonium states below the 
$D\bar{D}$ threshold.  
We keep the dependence on the factorization scale of the weak 
Hamiltonian $\mu$, but put the NRQCD factorization scale equal to 
$2 m_c$. At the scale $m_b=4.8\,$GeV we also have 
$\alpha_s(m_b)=0.22$ and $C_{[1]}(m_b)=0.55$, $C_{[8]}(m_b)=2.14$.
The tables also include the penguin correction factor 
$\delta_P$. 

We find that the dependence on the quark masses changes little 
when going from LO to NLO (for those $n$ for which the LO term is 
non-zero). The quark mass dependence is reasonably 
well estimated by that of the ratio
\begin{equation}
r = \frac{(1-4 m_c^2/m_b^2)^2}{m_b^2 m_c f_1(m_c/m_b)},
\end{equation}
where $f_1(z)$ is the tree-level phase space factor in round brackets 
in (\ref{semi}) and the NRQCD matrix elements are assumed fixed. 
If we vary $m_c$ by $\pm 100\,$MeV and $m_b$ by $\pm 200\,$MeV around 
our `standard' values and 
add the separate variations in square, we find a variation of $r$ of 
about 15\% relative to the standard value. Taking into account 
the approximativeness of this estimate, we assign an overall 
normalization uncertainty of 20\% due to quark masses. 

\subsubsection{$B\to \psi(nS)+X$}

\noindent 
We now turn to a more specific discussion of $B$ decay into the 
spin-triplet $S$-wave states $J/\psi$ and $\psi'$, denoted 
collectively by $\psi(nS)$ or just $\psi$. For quick reference 
we give the branching ratio in completely numerical form for 
$\mu=m_b=4.8\,$GeV and $m_c=1.5\,$GeV:\footnote{In this equation 
and those of the following ones that are similar in form, all numbers 
are given in units of GeV$^{-3}$.}
\begin{eqnarray}
\label{jpsitot}
\mbox{Br}\,(B\to \psi(nS)+X) &=& 
\left\{
\begin{array}{c}
-0.741 \\
0.0754 \\
-0.254 \\
\end{array}
\right\}
10^{-2}\,\langle {\cal O}^{\psi}_1(^3\!S_1) \rangle   +
0.195 \langle {\cal O}^{\psi}_8(^3\!S_1)  \rangle \\
&&\hspace*{-2cm}
+\,
0.342  \left[\langle {\cal O}^{\psi}_8(^1\!S_0)  \rangle  +
\frac{3.10}{m_c^2}\,\langle {\cal O}^{\psi}_8(^3\!P_0) \rangle
\right].
\nonumber
\end{eqnarray}
The penguin correction is included. For the coefficient of 
the colour singlet operator we
display the result obtained according to the procedures 
`NLO', `Improved' and `Total' in Table~\ref{tabsing}. As discussed 
earlier, we use the second entry (`Improved') in the following.

%%%%%%%%%%%%%%%%%%%%%%%%%%%%%%%%%%%%%%%%%%%%%%%%%%%%%%%%%%%%%%%%%%%
\begin{figure}[t]
   \vspace{-1.5cm}
   \epsfysize=16cm
   \epsfxsize=12cm
   \centerline{\epsffile{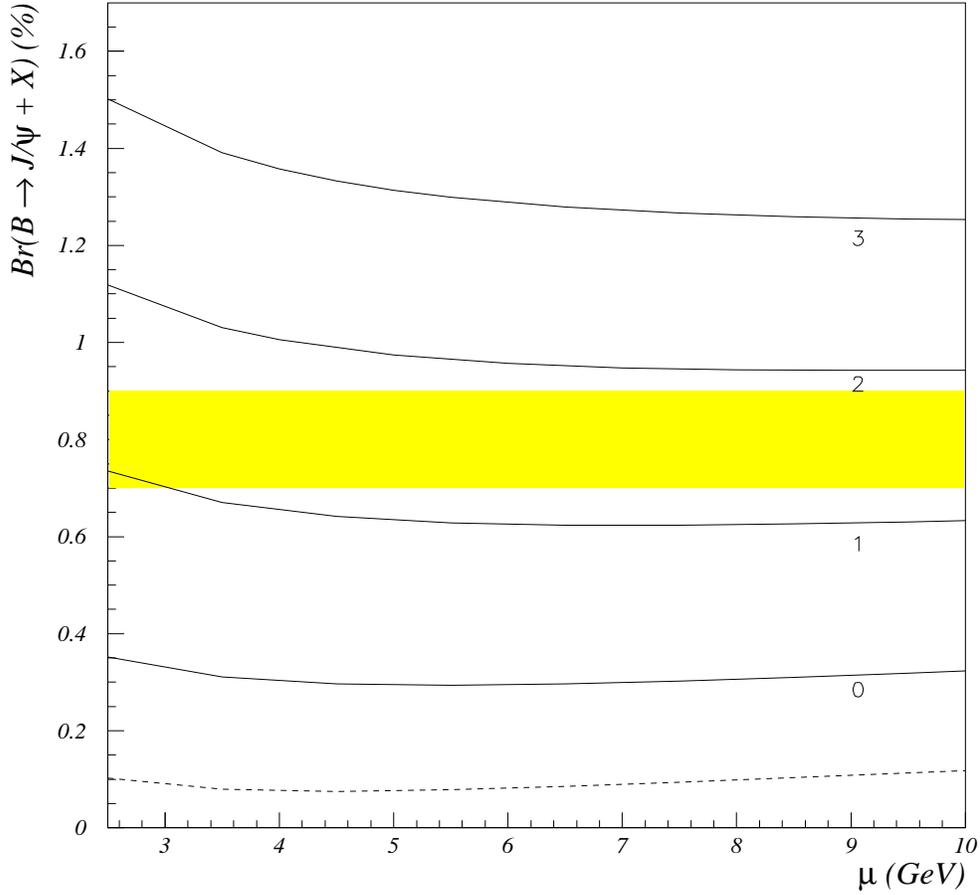}}
   \vspace*{-1.5cm}
\caption[dummy]{The $J/\psi$ branching fraction as a function of 
renormalization scale $\mu$ for various values of 
$M_{3.1}^\psi({}^1\!S_1^{(8)},{}^3\!P_J^{(8)})$ in $10^{-2}\,$GeV$^3$. 
The horizontal band shows the CLEO measurement \cite{CLEO95} and the 
dashed curve the colour singlet contribution alone. 
\label{figd}}
\end{figure}
%%%%%%%%%%%%%%%%%%%%%%%%%%%%%%%%%%%%%%%%%%%%%%%%%%%%%%%%%%%%%%%%%%%
The colour singlet matrix element is computed from the wave functions 
at the origin obtained with the Buchm\"uller-Tye potential as given 
in Ref.~\cite{EQ95}:
\begin{equation}
\langle {\cal O}_1^{\psi}({}^3\!S_1)\rangle=\frac{9|R(0)|^2}{2\pi}
=\left\{\begin{array}{c}
\,\,1.16\,\mbox{GeV}^3\qquad (J/\psi) \\
\!0.76\,\mbox{GeV}^3\qquad (\psi')
\end{array}
\right.
\end{equation}
The colour octet matrix element $\langle {\cal O}_8^{\psi}({}^3\!S_1)
\rangle$ is rather well determined by direct $\psi$ production at large 
transverse momentum in $p\bar{p}$ collisions \cite{BF95,CL96,BK97}. We 
use the values \cite{BK97}:
\begin{equation}
\langle {\cal O}_8^{\psi}({}^3\!S_1)\rangle=\left\{\begin{array}{c}
\,\,\,1.06\cdot 10^{-2}\,\mbox{GeV}^3\qquad (J/\psi) \\
\!0.44\cdot 10^{-2}\,\mbox{GeV}^3\qquad (\psi')
\end{array}
\right.
\end{equation}
There is an uncertainty of a factor 2 in each direction of the 
central value associated with these numbers.
With the number quoted the ${}^3\!S_1^{(8)}$ channel contributes 
0.21\% to the $J/\psi$ branching fraction and 0.09\% to the $\psi'$ 
branching fraction. The other two colour octet matrix elements are 
not yet well determined. In Fig.~\ref{figd} we show the 
$J/\psi$ branching fraction as a function of the  
renormalization scale $\mu$ for various values of 
$M_{3.1}^\psi({}^1\!S_1^{(8)},{}^3\!P_J^{(8)})$, where 
\begin{equation}
M_{k}^\psi({}^1\!S_1^{(8)},{}^3\!P_J^{(8)})= 
\langle {\cal O}^{\psi}_8(^1\!S_0)  \rangle  +
\frac{k}{m_c^2}\,\langle {\cal O}^{\psi}_8(^3\!P_0) \rangle.
\end{equation}
With the other parameters fixed, the branching ratios 
$\mbox{Br}\,(B\to J/\psi+X)=(0.80\pm 0.08)\%$ and 
$\mbox{Br}\,(B\to \psi'+X)=(0.34\pm 0.05)\%$ measured 
by CLEO are reproduced by
\begin{equation}
\label{central}
M_{3.1}^\psi({}^1\!S_1^{(8)},{}^3\!P_J^{(8)})
=\left\{\begin{array}{c}
\,\,\,1.5\cdot 10^{-2}\,\mbox{GeV}^3\qquad (J/\psi) \\
0.6\cdot 10^{-2}\,\mbox{GeV}^3\qquad (\psi').
\end{array}
\right.
\end{equation}
If we allow the colour singlet contribution to vary between zero 
and twice the value assumed above, include the above variation 
of $\langle {\cal O}_8^{\psi}({}^3\!S_1)\rangle$ as well as the experimental 
error, and add all variations linearly, we obtain the allowed range
\begin{equation}
\label{range}
M_{3.1}^\psi({}^1\!S_1^{(8)},{}^3\!P_J^{(8)})
=\left\{\begin{array}{c}
\,\,\,(0.4-2.3)\cdot 10^{-2}\,\mbox{GeV}^3\qquad (J/\psi) \\
(0.0-1.0)\cdot 10^{-2}\,\mbox{GeV}^3\qquad (\psi').
\end{array}
\right.
\end{equation}
It is interesting to compare the central values (\ref{central}) 
and the upper limits with other determinations of the parameter 
$M_{k}^\psi({}^1\!S_1^{(8)},{}^3\!P_J^{(8)})$. The central values 
are about a factor 3 smaller than the central values obtained 
for $M_{3.5}^\psi({}^1\!S_1^{(8)},{}^3\!P_J^{(8)})$
from $\psi$ production at the Tevatron at moderate transverse 
momentum \cite{CL96,BK97}. As emphasized in Ref.~\cite{BK97} the 
Tevatron extraction is very sensitive to various effects that 
affect the transverse momentum distribution. Indeed, 
Refs.~\cite{CC97,KK98} quote smaller values compatible with, 
or smaller than the central values above. The total production 
cross section in fixed target collisions probes 
$M_{7}^\psi({}^1\!S_1^{(8)},{}^3\!P_J^{(8)})$ (assuming the 
validity of NRQCD factorization, which may be controversial). 
Given that a different combination of matrix elements enters, 
the values obtained in Ref.~\cite{BR96} are certainly consistent with 
the above central value. In view of the uncertainties involved in 
charmonium production in hadron collisions, we believe that the 
above upper limit on $M_{3.1}^\psi({}^1\!S_1^{(8)},{}^3\!P_J^{(8)})$ 
is the most stringent one existing at present. We note that 
small values of $\langle {\cal O}^{\psi}_8(^1\!S_0) \rangle$ and 
$\langle {\cal O}^{\psi}_8(^3\!P_0) \rangle$ seem to be preferred 
by the non-observation of a significant colour octet contribution 
in the energy spectrum of inelastic $J/\psi$ photoproduction 
\cite{CK96,KLS96,BKV98} (see, however, 
the discussion in Ref.~\cite{BRW97}).
We conclude that the measured $J/\psi$ and $\psi'$ branching fractions 
can be accounted for with values of the NRQCD long-distance 
parameters consistent with previously available values. 

\subsubsection{$B\to \eta_c+X$}

\noindent
Presently, only an experimental 
upper bound $\mbox{Br}\,(B\to \eta_c+X)<0.9\%$ 
\cite{CLEO95} exists on $\eta_c$ production. For the same choice 
of input parameters as above, we have
\begin{eqnarray}
\mbox{Br}\,(B\to \eta_c+X) &=& 
\left\{
\begin{array}{c}
-1.19 \\
0.250 \\
-0.210 \\
\end{array}
\right\}
10^{-2}\,\langle {\cal O}^{\eta_c}_1(^1\!S_0) \rangle   +
0.342 \langle {\cal O}^{\eta_c}_8(^1\!S_0)  \rangle \\
&&\hspace*{-2cm}
+\,
0.195  \left[\langle {\cal O}^{\eta_c}_8(^3\!S_1)  \rangle  -
\frac{0.240}{m_c^2}\,\langle {\cal O}^{\eta_c}_8(^1\!P_1) \rangle
\right].
\nonumber
\end{eqnarray}
The LO term is enhanced by about 10\% because of the penguin correction. 

There is at present 
no information on the $\eta_c$ colour octet matrix elements 
from other $\eta_c$ production processes. The colour octet matrix 
elements are non-zero because soft gluon emission connects the colour 
octet $c\bar{c}$ state to the physical charmonium state. The soft gluon 
emission amplitude can be multipole expanded, supposing  
that the characteristic momentum of the emitted gluons is of order 
$m_c v^2$, smaller than the characteristic momentum $m_c v$ of the charm 
quarks in the charmonium rest frame. Up to corrections of order 
$v^2$, spin symmetry imposes relations between the $\eta_c$ and 
$J/\psi$ matrix elements. In addition to the familiar spin symmetry 
relation
$\langle {\cal O}^{J/\psi}_1(^3\!S_1)\rangle 
=3\,\langle {\cal O}^{\eta_c}_1(^1\!S_0)\rangle$ for the 
colour singlet wave function, we find
\begin{eqnarray}
\label{rels}
\langle {\cal O}^{\eta_c}_8(^1\!S_0)\rangle &=& 
\frac{1}{3}\,\langle {\cal O}^{J/\psi}_8(^3\!S_1)\rangle,
\nonumber\\
\langle {\cal O}^{\eta_c}_8(^3\!S_1)\rangle &=& 
\langle {\cal O}^{J/\psi}_8(^1\!S_0)\rangle,
\\
\langle {\cal O}^{\eta_c}_8(^1\!P_1)\rangle &=& 
3\,\langle {\cal O}^{J/\psi}_8(^3\!P_0)\rangle.
\nonumber
\end{eqnarray}
Note that these relations are consistent with the renormalization group 
equations for the matrix elements that follow from (\ref{rgeir}). 
Since we do not know $\langle {\cal O}^{J/\psi}_8(^1\!S_0)\rangle$ and 
$\langle {\cal O}^{J/\psi}_8(^3\!P_0)\rangle$ separately, we assume that 
between one half and all of 
$M_{3.1}^\psi({}^1\!S_1^{(8)},{}^3\!P_J^{(8)})$ is due 
to $\langle {\cal O}^{J/\psi}_8(^1\!S_0)\rangle$. With the 
central value from (\ref{central}) this leads to 
the estimate
\begin{equation}
\mbox{Br}\,(B\to \eta_c+X) \approx (0.3-0.5)\%.
\end{equation}
We emphasize that this estimate is crude and depends sensitively 
on the validity of the relations (\ref{rels}). This estimate is 
below the $J/\psi$ branching 
fraction, but with the increase in statistics since the previous analysis 
\cite{CLEO95}, a branching fraction in the above range may perhaps be 
reached with the CLEO detector.

\subsubsection{$B\to \chi_{cJ}+X$}

\noindent
Colour octet effects in charmonium production were in fact considered 
for the first time for $\chi_c$ production in $B$ decay \cite{BBYL92}. 
The authors showed that the observed $\chi_{c2}$ signal can be explained 
by the production of a $c\bar{c}[{}^3\!S_1^{(8)}]$ state followed 
by a soft dipole transition. (Recall that at LO in the colour singlet 
model, $\chi_{c0}$ and $\chi_{c2}$ are not produced.) The LO production 
through a $c\bar{c}[{}^3\!S_1^{(8)}]$ pair corresponds to the IR 
sensitive contribution at order $\alpha_s$ in the ordinary colour 
singlet channel. Our NLO calculation adds to those the `hard'
contributions at order $\alpha_s$ in the colour singlet and the 
colour octet channels. With $m_b=4.8\,$GeV, $m_c=1.5\,$GeV and 
$\mu=m_b$ as usual we obtain
\begin{eqnarray}
\mbox{Br}\,(B\to\chi_{c0}+X) &=& 
\frac{-0.0148}{m_c^2}\,\langle {\cal O}^{\chi_{c0}}_1(^3\!P_0)\rangle  +
0.195 \,\langle {\cal O}^{\chi_{c0}}_8(^3\!S_1)  \rangle,
\\
\label{chi1}
\mbox{Br}\,(B\to\chi_{c1}+X) &=& 
\left\{
\begin{array}{c}
-2.14\\
-0.783 \\
-1.21 \\
\end{array}
\right\}
\frac{10^{-2}}{m_c^2}\,\langle {\cal O}^{\chi_{c1}}_1(^3\!P_1)\rangle  +
0.195  \,\langle {\cal O}^{\chi_{c1}}_8(^3\!S_1)  \rangle,
\\
\label{chi2}
\mbox{Br}\,(B\to\chi_{c2}+X) &=& 
\frac{-0.0120}{m_c^2}\,\langle {\cal O}^{\chi_{c2}}_1(^3\!P_2)\rangle  +
0.195  \,\langle {\cal O}^{\chi_{c2}}_8(^3\!S_1)  \rangle
\end{eqnarray}
to be compared with the measurements \cite{CLEO95}
\begin{eqnarray}
\mbox{Br}\,(B\to\chi_{c1}+X) &=& (0.37\pm 0.07)\%,
\\
\mbox{Br}\,(B\to\chi_{c2}+X) &=& (0.25\pm 0.10)\%.
\end{eqnarray}
Owing to the spin symmetry relations, valid up to higher order corrections 
in $v^2$,
\begin{eqnarray}
\langle {\cal O}^{\chi_{cJ}}_1(^3\!P_J)\rangle 
&=& (2J+1)\,\langle {\cal O}^{\chi_{c0}}_1(^3\!P_0)\rangle,
\\
\langle {\cal O}^{\chi_{cJ}}_8(^3\!S_1) \rangle
&=& (2J+1)\,\langle {\cal O}^{\chi_{c0}}_8(^3\!S_1)\rangle,
\end{eqnarray}
only two of the six above parameters are independent. The colour singlet 
contribution is always negative. However, since the two matrix elements 
involved mix under renormalization each short-distance coefficient 
depends on an arbitrary convention to separate the two contributions. 
(We used the $\overline{\rm MS}$ scheme.) Hence a negative partial rate 
is not unphysical.

Due to the near-proportionality of (\ref{chi1}), (\ref{chi2}), we find,  
however,  
that it is very difficult to reproduce the experimental result with a 
reasonable choice of matrix elements. With most reasonable choices one 
obtains a $\chi_{c2}$ production cross section larger than the cross 
section for $\chi_{c1}$. If we take the colour singlet matrix 
element from the Buchm\"uller-Tye potential model \cite{EQ95}, 
\begin{equation}
\label{m1}
\langle {\cal O}^{\chi_{c0}}_1(^3\!P_0)\rangle/m_c^2 = 4.8\cdot 10^{-2}\,
\mbox{GeV}^3,
\end{equation}
and adjust\footnote{This value can be compared with 
$\langle {\cal O}^{\chi_{c0}}_8(^3\!S_1)\rangle  = 3.2\cdot 10^{-3}\,
\mbox{GeV}^3$ obtained in 
Ref.~\cite{CL96} from 
$\chi_{cJ}$ production at the Tevatron collider.}
\begin{equation}
\label{m2}
\langle {\cal O}^{\chi_{c0}}_8(^3\!S_1)\rangle  = (4.5-6.5)\cdot 10^{-3}\,
\mbox{GeV}^3
\end{equation}
to reproduce the measured $\chi_{c2}$ branching fraction, we obtain
\begin{equation}
\mbox{Br}\,(B\to\chi_{c1}+X) = (0.15-0.27)\%,
\end{equation}
which is below the measurement. 
We conclude that the expansion in $\alpha_s$ is not well behaved 
enough for $P$-wave charmonium production in $B$ decay, at least 
to next-to-leading order, 
to arrive at quantitative relations. The problem is caused to a large 
extent by the fact that the leading order colour singlet contribution 
to $\chi_{c1}$ production, which would have been expected to enhance 
$\chi_{c1}$ production relative to $\chi_{c2}$ production, is turned 
negative (or very small) at NLO. 

In addition, the colour singlet contributions are also negative for 
$\chi_{c0}$ and $\chi_{c2}$, which have no leading order contribution. 
This requires large cancellations between a negative colour 
singlet contribution and a positive colour 
octet contribution. These cancellations may be considered an 
artefact of the $\overline{\rm MS}$ factorization scheme, which 
appears unnatural from this point of view.

Because of this unsatisfactory situation, 
we find it difficult to predict 
the $\chi_{c0}$ branching fraction better than the naive expectation of 
one fifth of the $\chi_{c2}$ branching fraction. Note that we consider 
the prediction for the $\chi_{c1}$ state less reliable than for 
the $\chi_{c2}$ state, so that the range for the colour octet matrix 
element obtained in (\ref{m2}) may not be totally unreasonable.

\subsubsection{$B\to h_c+X$}

\noindent
Finally, we consider the production of the ${}^1\!P_1$ state 
$h_c$, which will probably be difficult to measure. With the usual 
choice of parameters:
\begin{eqnarray}
\mbox{Br}\,(B\to h_c+X) &=& 
\frac{-0.0228}{m_c^2}\,\langle {\cal O}^{h_c}_1(^1\!P_1)\rangle  +
0.342\,\langle {\cal O}^{h_c}_8(^1\!S_0)  \rangle.
\end{eqnarray}
We use the 
spin symmetry relation $\langle {\cal O}^{h_c}_1(^1\!P_1)\rangle 
=3\,\langle{\cal O}^{\chi_{c0}}_1(^3\!P_0)\rangle$ for the derivative 
of the wave function at the origin (squared) and the spin symmetry 
relation 
\begin{equation}
\label{est}
\langle {\cal O}^{h_c}_8(^1\!S_0)\rangle 
=3\,\langle{\cal O}^{\chi_{c0}}_8(^3\!S_1)\rangle
\end{equation}
for the colour octet matrix element. 
With the value of the colour singlet matrix element as quoted above 
and the colour octet matrix element in the range (\ref{m2}) 
we obtain the estimate
\begin{equation}
\mbox{Br}\,(B\to h_c+X) \approx (0.13-0.34)\%.
\end{equation}
The given range hinges on the estimate (\ref{est}), which   
follows from the statistical factor 3 for the E1 ${}^1\!S_0\to {}^1\!P_1$ 
transition and another factor 3 for the final state with $J=1$, 
provided we assume a spin-independent overlap integral in terms of 
colour singlet and colour octet wave functions. Then taking into account 
the fact 
that the definition of $\langle{\cal O}^{\chi_{c0}}_8(^3\!S_1)\rangle$ 
does not include an average over the three initial polarizations, we obtain 
the relative factor $9/3=3$. Note again that (\ref{est}) is consistent 
with (\ref{rgeir}).

\subsection{Quarkonium momentum distributions}

\noindent 
We briefly address quarkonium momentum distributions. We restrict 
ourselves to the $S$-wave states $J/\psi$ and $\psi'$, first because 
data exist at present only for these states \cite{CLEO95} and 
second, because the theoretical prediction appears to be most 
reliable for the $S$-wave states.

The energy distributions of the $c\bar{c}$ -pair in the partonic 
process $b\to c\bar{c}[n]+q+g$ are collected in Appendix B. The 
partonic energy/momentum  distributions are distributions 
in the mathematical sense and cannot be used to predict the physical 
momentum spectrum. Two effects lead to a smearing of the partonic 
energy/momentum distribution:

(a) The $b$ quark has a residual motion in the $B$ meson and does 
not decay at rest. This leads to an energy smearing of order 
$\Lambda$, where $\Lambda$ is the QCD scale. 
For $B$ decay into charmonium this `Fermi motion' 
effect has already been modelled in Ref.~\cite{BKLP81}, and more 
recently in Ref.~\cite{PPS97}.

(b) The second effect is related to the fact that charmonium 
production through colour octet $c\bar{c}[n]$ states requires the 
emission of soft gluons with energy of order $m_c v^2\approx 
\Lambda$. The NRQCD matrix elements measure the probability for 
the transition $n\to\,\mbox{charmonium}$, but do not take into 
account the kinematic effect of the soft gluon emission, which 
becomes important near the upper endpoint of the charmonium 
momentum distribution \cite{BRW97}. Because the maximal $J/\psi$ 
momentum in $B$ decay is only about $2\,$GeV, this effect is 
expected to be at least as important as the $b$ quark motion, especially 
in view of the large fraction of colour octet production. No attempt 
has been made so far to take this effect into account in a realistic 
model for the spectrum.
Both effects are also related to the fact that the 
phase space boundaries in the partonic calculation depend on the 
quark masses rather than the $B$ meson, $J/\psi$ and 
$K$ masses.

\begin{table}[t]
\addtolength{\arraycolsep}{0.25cm}
\renewcommand{\arraystretch}{1.6}
$$
\begin{array}{c|c|c|c|c}
\hline\hline
\mbox{Moment} &  \langle {\cal O}^{\psi}_1(^3\!S_1) \rangle & 
\langle {\cal O}^{\psi}_8(^3\!S_1)  \rangle & 
\langle {\cal O}^{\psi}_8(^1\!S_0)  \rangle & 
\langle {\cal O}^{\psi}_8(^3\!P_0) \rangle/m_c^2 \\ 
\hline
0         & 7.5\cdot 10^{-4} & 2.0\cdot 10^{-1} & 3.4\cdot 10^{-1} & 
1.06   \\
\mbox{L1} & 5.1\cdot 10^{-4} & 1.9\cdot 10^{-1} & 3.3\cdot 10^{-1} & 
9.8\cdot 10^{-1}   \\
\mbox{L2} & 3.3\cdot 10^{-4} & 1.8\cdot 10^{-1} & 3.2\cdot 10^{-1} & 
9.2\cdot 10^{-1}   \\
\mbox{L3} & 1.7\cdot 10^{-4} & 1.7\cdot 10^{-1} & 3.1\cdot 10^{-1} & 
8.7\cdot 10^{-1}   \\
\hline
\mbox{S1} & 2.4\cdot 10^{-4} & 8.7\cdot 10^{-3} & 1.4\cdot 10^{-2} & 
7.9\cdot 10^{-2}   \\
\mbox{S2} & 5.4\cdot 10^{-5} & 1.6\cdot 10^{-3} & 2.3\cdot 10^{-3} & 
1.6\cdot 10^{-2}   \\
\mbox{S3} & 2.4\cdot 10^{-5} & 5.8\cdot 10^{-4} & 7.5\cdot 10^{-4} & 
6.1\cdot 10^{-3}   \\
\hline\hline
\end{array}
$$
\caption[dummy]{\label{tabmom} Coefficients of NRQCD matrix elements 
in GeV$^{-3}$ for the 
moments of the $J/\psi$ momentum distribution. `Ln' denotes the 
moment with the weight function $z^n$, `Sn' the moment with 
the weight function $(1-z)^n$.}
\end{table}

In this paper neither of the two effects will be modelled. One 
expects them to be suppressed by powers of $\Lambda_{QCD}/m_{b,c}$ and 
$v^2$ provided 
a smooth average of the momentum distribution (like the integration 
to the total width) is taken. We define
\begin{eqnarray}
\label{ws}
\mbox{Br}\,(\psi,W) &\equiv& 
\int\limits_0^1 dz \,W(z)\,\frac{1}{\Gamma_B}\,
\frac{d\Gamma(B\to \psi+X)}{dz}
\nonumber\\
&=&\sum_n {\cal N}\,
\langle {\cal O}^{\psi}[n]\rangle \int\limits_{2\sqrt{\eta}}^{1+\eta} 
dx\,\frac{d\Gamma[n]}{dx}\,W\!\left(\frac{\sqrt{x^2-4\eta}}{1-\eta}\right),
\end{eqnarray}
where $z\equiv |\vec{p}_{\psi}|/|\vec{p}_{\psi,max}|$ and  $\Gamma_B$ is 
the total $B$ decay rate. The variable $x$ is the energy 
fraction, defined together with $d\Gamma[n]/dx$ in (\ref{notedist}). 
(The functions $g_i[n](\eta,x)$ that enter there can be found in 
Appendix~B.) Finally $W(z)$ is a smooth weight on the momentum 
distribution. 

In Table~\ref{tabmom} we give the coefficients in front of the 
NRQCD matrix elements in (\ref{ws}) for the weight functions 
$z^n$ (Ln) and $(1-z)^n$ (Sn) up to $n=3$. For $n=0$ we recover 
the inclusive branching fraction (\ref{jpsitot}).\footnote{The precise 
implementation is done as follows: For the delta-function term in 
(\ref{red}) we use the `improved' prescription for the colour 
singlet channel. For the second term on the right-hand side of 
(\ref{red}) we use the strict NLO approximation. This is reasonable, 
because the negative contributions that necessitate the 
improvement are mainly associated with virtual corrections.} 
The first weight function increasingly weights the endpoint region 
as $n$ increases. Therefore one cannot take $n$ large without 
enhancing higher order terms in the velocity expansion \cite{BRW97} 
not taken into account here. The second weight function weights 
the small-momentum tail. The moments decrease rapidly with $n$, because, 
as expected, the spectrum favours a large momentum of the $\psi$. 
The LO order and NLO virtual contributions do not contribute 
to the Sn moments. These moments are directly sensitive to hard gluon 
radiation.

The momentum spectrum measured by CLEO \cite{CLEO95} is given in the 
CLEO rest frame rather than the $B$ rest frame. With the improved 
statistics that should be available now, it will be interesting 
to see whether one can obtain additional information on charmonium 
production by comparing averages of the momentum spectrum with the 
above predictions. For example, one may think of using the Sn moments 
to determine the parameters $\langle {\cal O}^{\psi}_8(^1\!S_0) \rangle$ 
and $\langle {\cal O}^{\psi}_8(^3\!P_0) \rangle$ separately.  

\noindent
\subsection{Comparison with exclusive two-body modes}

\noindent
It is interesting to compare qualitative features of the theoretical 
result for inclusive charmonium production with the sum of the 
most important exclusive two-body decay channels containing charmonium. 
We discuss only $J/\psi$ because only limited experimental information 
exists for the other charmonium states.

There exist only two two-body modes of any importance. Their branching 
fractions are measured to be \cite{PDG}
\begin{eqnarray}
\label{bb1}
\mbox{Br}\,(B\to J/\psi+K) &=& \left\{\begin{array}{c}
(0.099\pm 0.010)\%\\
(0.089\pm 0.012)\%,\end{array}
\right.
\\
\label{bb2}
\mbox{Br}\,(B\to J/\psi+K^*) &=& \left\{\begin{array}{c}
(0.147\pm 0.027)\%\\
(0.135\pm 0.018)\%,\end{array}
\right.
\end{eqnarray}
where the upper line in both (\ref{bb1}) and (\ref{bb2}) refers to 
$B^{\pm}$ and the lower one 
to $B^0$. The combined branching fraction of about 0.25\% is far below 
the inclusive branching fraction of $0.80\pm 0.08\%$. The existence of 
a large fraction of three-body modes is also confirmed by the 
broad energy distribution measured by CLEO \cite{CLEO95}.

There is no clear association of the colour singlet contribution with the 
inclusive branching fraction with the above two-body modes. A colour 
singlet $c\bar{c}$ pair can radiate a hard gluon and end up as a 
three-body mode. Likewise, the soft gluon that is necessarily emitted 
from a $c\bar{c}$ pair in a colour octet state can be re-absorbed 
by the light quarks and combine to the kaon. This process depends on the 
spectator light quark in the $B$ meson and would therefore seem to 
violate the factorization hypothesis of the NRQCD approach. However, 
factorization requires only that such spectator dependence cancels out 
to first approximation (in $\Lambda_{QCD}/m_{c,b}$) in the 
average over all decay modes. 

Nevertheless it is natural to expect that a $c\bar{c}$ pair in a 
colour octet state finds itself more often hadronizing in a multi-body 
decay than a colour singlet $c\bar{c}$ pair and therefore we consider 
the observation of a large fraction of multi-body decays as supporting 
evidence for the colour octet picture. The fact that the sum of 
the two two-body modes above is larger than the (poorly predicted) 
inclusive colour singlet branching fraction of approximately 0.1\% 
suggests that the NLO calculation underestimates this contribution. 
(Another possibility is that some fraction of the large colour octet 
partial rate does in fact end up in two-body modes.)

The large fraction of three body decays is significant in another 
respect. The NRQCD approach assumes `local parton-hadron 
duality', which is often discussed in connection with the 
inclusive non-leptonic decays of $B$ mesons. A crucial consequence 
of `local parton-hadron duality' is that the effect of colour 
reconnections to the spectator quarks are small (power-suppressed) 
even though colour reconnections must occur in every single decay. 
The same assumption underlies the NRQCD factorization approach to 
inclusive charmonium production. The assumption is usually justified 
by the presence of a large energy release into the final state and 
many decay modes to be averaged over. Since the energy release in 
decays into charmonium is not particularly large, the existence of 
a sufficient fraction of decays with additional pions suggests 
that the total decay rate provides enough averaging for an 
approximate cancellation of non-factorizable effects.\footnote{We 
use the term `non-factorizable' in the sense of NRQCD factorization. 
In the literature on two-body decays of $B$ mesons `non-factorizable' 
usually refers to any virtual gluon correction (below the scale 
$m_b$) that connects the 
$b$ or $s$ quark to the $c$ or $\bar{c}$ quark. These `non-factorizable' 
contributions are included in the inclusive NRQCD calculation, 
see Fig.~\ref{figa}.}

\section{Conclusion}
\label{conclusion}

\noindent 
We presented an analysis of inclusive $B$ decay into the known 
charmonium states at next-to-leading order in the strong coupling 
and accounting for the most important colour singlet and 
colour octet production mechanisms. 
We find that radiative corrections make the colour singlet 
contributions negative, an effect already observed in Ref.~\cite{BE94} 
for the colour singlet contribution to $J/\psi$ production and 
explained by the suppression of colour singlet production at 
leading order due to the particular structure of the weak 
effective Hamiltonian. The problem is particularly serious for 
$\chi_{c1}$ production. As a consequence, the appealing 
theoretical pattern for $\chi_{cJ}$ production at leading order 
\cite{BBYL92} becomes obscured quantitatively at next-to-leading 
order, although the qualitative requirement of an additional 
colour octet component is not put into question. In general, we 
find it difficult to make a quantitative prediction for any of 
the four $P$-wave states.

The situation is more satisfactory for $S$-wave production, and in 
particular for $J/\psi$ production, for which we confirm the earlier 
conclusion that the colour singlet contribution is about a factor 
5--10 below the observed production rate. The next-to-leading order 
corrections to the colour octet channels computed in this paper 
are positive and of order 20\%--50\%. Assuming that these channels 
make up for the missing contribution, we adjust a certain combination 
of NRQCD matrix elements to reproduce the experimental branching 
fraction. The value for the matrix element is somewhat smaller than, 
but within errors compatible with the magnitude suggested by other 
quarkonium production processes. Since the $\alpha_s$ expansion 
appears to be well-behaved for $J/\psi$ production, we obtain 
the sharpest upper bound to date on a combination of 
$\langle {\cal O}^{\psi}_8(^1\!S_0)  \rangle$ and 
$\langle {\cal O}^{\psi}_8(^3\!P_0) \rangle/m_c^2$, even when we 
assume that there are no other production mechanisms. In principle, 
with more accurate data appearing, weights on the $J/\psi$ momentum 
distribution can also be used to determine 
$\langle {\cal O}^{\psi}_8(^1\!S_0)  \rangle$ and 
$\langle {\cal O}^{\psi}_8(^3\!P_0) \rangle/m_c^2$ individually, 
which has so far proven to be difficult, even combining information 
from other $J/\psi$ production processes.\\

\noindent {\em Acknowledgements.} We thank Gerhard Buchalla and 
Marco Ciuchini for helpful discussions on the renormalization of the weak 
effective Hamiltonian in the HV scheme. We thank Ben Grinstein for 
his interest in this work and Tom Ferguson and Giancarlo Moneti for their 
effort in explaining the data on $J/\psi$ momentum distributions to us. 
This work was supported in part by the EU Fourth Framework Programme
`Training and Mobility of Researchers', Network `Quantum Chromodynamics and
the Deep Structure of Elementary Particles', contract FMRX-CT98-0194 (DG 12
- MIHT) and by the DOE grant DOE-ER-40682-145. 

\section*{Appendix A. Short-distance coefficients for integrated 
decay widths}
\renewcommand{\theequation}{A.\arabic{equation}}

\noindent
We collect the expressions that enter the integrated partial rates 
(\ref{notdecay}) in this appendix. Recall $\eta=4 m_c^2/m_b^2$. The scale 
$\mu$ denotes the factorization scale at which the coefficients 
$C_{[1,8]}$ are evaluated. We distinguish from it the renormalization 
scale $\hat{\mu}$ of the NRQCD matrix element $\langle {\cal O}^H[n]
\rangle$.

\subsection*{A.1. Lowest order functions}

\noindent
The lowest order function $f[n](\eta)$ vanishes for 
$n={}^3\!P_0^{(1)},\,^3\!P_2^{(1)},\,^1\!P_1^{(1)}$ and 
$^1\!P_1^{(8)}$. The non-vanishing ones are:

\begin{eqnarray}
f[^3\!S_1^{(1)}](\eta) &=& (1-\eta)^2 (1+2 \eta),
\\
f[^1\!S_0^{(1)}](\eta) &=& 3\, (1-\eta)^2,
\\
f[^3\!P_1^{(1)}](\eta) &=& 2\, (1-\eta)^2 (1+2 \eta),
\\
f[^3\!S_1^{(8)}](\eta) &=& \frac{3}{2}\,(1-\eta)^2 (1+2 \eta),
\\
f[^1\!S_0^{(8)}](\eta) &=& \frac{9}{2}\, (1-\eta)^2,
\\
f[^3\!P_J^{(8)}](\eta) &=& 9\, (1-\eta)^2 (1+2 \eta).
\end{eqnarray}

\subsection*{A.2. The penguin correction}

\noindent
We consider only the contribution where a decay through a QCD 
penguin operator ${\cal O}_{3-6}$ interferes with the decay amplitude 
through a current-current operator. Since the penguin operators 
have small coefficient functions, it is sufficient to evaluate the 
penguin contributions in lowest orders in $\alpha_s$. 
The double penguin contribution is negligible in size. 
Since $f[n](\eta)=0$ for 
$n={}^3\!P_0^{(1)},\,^3\!P_2^{(1)},\,^1\!P_1^{(1)},\,^1\!P_1^{(8)}$, 
the penguin contribution also vanishes in leading order for 
these $n$. For the 
other intermediate states, we find
\begin{equation}
\delta_P[^3\!S_1^{(1)}] = 2\,\frac{3 (C_3+C_5)+C_4+C_6}{C_{[1]}} 
\approx -0.004,
\end{equation}
\begin{equation}
\delta_P[^1\!S_0^{(1)}] = \delta_P[^3\!P_1^{(1)}] = 
2\,\frac{3 (C_3-C_5)+C_4-C_6}{C_{[1]}}
\approx 0.07,
\end{equation}
\begin{equation}
\delta_P[^3\!S_1^{(8)}] = 4\,\frac{C_4+C_6}{C_{[8]}}
\approx -0.09,
\end{equation}
\begin{equation}
\delta_P[^1\!S_0^{(8)}] = \delta_P[^3\!P_J^{(8)}] = 
4\,\frac{C_4-C_6}{C_{[8]}}
\approx 0.009.
\end{equation}
Here we used unitarity of the Cabibbo-Kobayashi-Maskawa (CKM) 
matrix to relate the CKM factor of the penguin contribution to the CKM 
factor of the current-current operator contribution (up to a negligible 
error $\mbox{Re}\,(V_{cb}^* V_{cs}/(V_{tb}^* V_{ts}))=-1$). 
For the numerical estimate we have used the leading logarithmic 
approximation for the Wilson coefficients at the scale $m_b=4.8\,$GeV  
and with $\Lambda_{QCD}^{LO}$ adjusted to reproduce $\alpha_s(m_Z)$ 
which gives $\Lambda_{QCD}^{LO}=93\,$MeV (for 5 flavours). This implies 
$C_{[1]}(m_b)=0.41$, $C_{[8]}(m_b)=2.19$, $C_3(m_b)=0.010$,
$C_4(m_b)=-0.024$,  $C_5(m_b)=0.007$ and $C_6(m_b)=-0.028$. 

\subsection*{A.3. Next-to-leading order functions}

\noindent
The next-to-leading order functions depend on the scheme-dependent 
constants $X_R$, $Y_R$ and $Z_R$. Their values in the NDR and HV scheme 
are given in section~\ref{uvreg}. 

\subsubsection*{A.3.1. $^3\!S_1^{(1)}$}

\noindent
\begin{eqnarray}
g_1[^3\!S_1^{(1)}](\eta) &=& 
\frac{4}{3}\Bigg\{(-1)\,(1-\eta)^2 (1+2\eta)
\bigg[8 \,\mbox{Li}_2(\eta)+4\ln(1-\eta)\ln\eta-4 Z_R+\frac{4\pi^2}{3}
\bigg]\nonumber\\
&&\hspace*{-1.5cm}-\,4 \,\eta (1+\eta) \,(1-2\eta)\,\ln\eta - 
2 \,(1-\eta)^2 \,(5+4\eta) \,\ln(1-\eta)\nonumber\\
&&\hspace*{-1.5cm}-\,(1-\eta) \,(1-2\eta) \,(3+5\eta)\Bigg\}.
\\[0.2cm]
g_2[^3\!S_1^{(1)}](\eta) &=& 
\frac{4}{3}\Bigg\{(1-\eta)^2 (1+2\eta)\left[3\ln\frac{m_b^2}{\mu^2}
+Y_R-X_R\right]\nonumber\\
&&\hspace*{-1.5cm}-\,
\frac{(1-\eta) (34+23\eta
-51\eta^2+16\eta^3)}{2 (2-\eta)}
+\frac{2(1-\eta)^3 (3-\eta^2)}{(2-\eta)^2}
\ln(1-\eta)\nonumber\\
&&\hspace*{-1.5cm}-\,\frac{(26-19\eta+4\eta^2)\eta^2}{2-\eta}\ln\eta
-\frac{8(1-\eta)^3\eta}{2-\eta}\ln 2\Bigg\}.
\\[0.2cm]
g_3[^3\!S_1^{(1)}](\eta) &=& 
\frac{4}{27}\,(1-\eta) \,(1+37\eta-8\eta^2) - \frac{8\,(1-6\eta)}{9}\,
\ln\eta.
\end{eqnarray}

\subsubsection*{A.3.2. $^1\!S_0^{(1)}$}

\noindent
\begin{eqnarray}
g_1[^1\!S_0^{(1)}](\eta) &=& 
(-4)\,(1-\eta)^2 
\bigg[8 \,\mbox{Li}_2(\eta)+4\ln(1-\eta)\ln\eta-4 Z_R+\frac{4\pi^2}{3}
\bigg]\nonumber\\
&&\hspace*{-1.5cm}-\,16 \,\eta (1-\eta) \,\ln\eta + 
\frac{8\,(1-\eta)^2 \,(2-5\eta)}{\eta} \,\ln(1-\eta) 
+ 20\,(1-\eta)^2.\\[0.2cm]
g_2[^1\!S_0^{(1)}](\eta) &=& 
 4 (1-\eta)^2 \left[3\ln\frac{m_b^2}{\mu^2}
+Y_R-X_R\right] + 4\eta^2\,\ln\eta 
\nonumber\\
&&\hspace*{-1.5cm}
-\,\frac{2 (1-\eta) (34-53\eta+17\eta^2)}{2-\eta}
+\frac{8\,(1-\eta)^3 (3-\eta)}{(2-\eta)^2}\,\ln(1-\eta).\\[0.2cm]
g_3[^1\!S_0^{(1)}](\eta) &=& 
\frac{4}{9}\,\left((-1)\,(1-\eta)\,(11-7\eta+2\eta^2) - 6\ln\eta\right).
\end{eqnarray}

\subsubsection*{A.3.3. $^3\!P_0^{(1)}$}

\noindent
\begin{eqnarray}
g_1[^3\!P_0^{(1)}](\eta) &=& 0.\\[0.2cm]
g_2[^3\!P_0^{(1)}](\eta) &=& 0.\\[0.2cm]
g_3[^3\!P_0^{(1)}](\eta) &=& 
 \frac{16}{9}\,(1-\eta)^2 \,(1+2 \eta)\,\left[-\ln\frac{\hat{\mu}^2}{4 m_c^2}
+ 2 \ln(1-\eta)\right]\nonumber\\
&&\hspace*{-1.5cm}
-\,\frac{8}{9}\,(1-12\eta^2+8\eta^3)\,\ln\eta - 
 \frac{4}{9}\,(1-\eta)\,(25-13\eta-18\eta^2).
\end{eqnarray}

\subsubsection*{A.3.4. $^3\!P_1^{(1)}$}

\noindent
\begin{eqnarray}
g_1[^3\!P_1^{(1)}](\eta) &=& 
\left(-\frac{8}{3}\right)\,(1-\eta)^2 \,(1+2\eta)
\bigg[8 \,\mbox{Li}_2(\eta)+4\ln(1-\eta)\ln\eta-4 Z_R+\frac{4\pi^2}{3}
\bigg]\nonumber\\
&&\hspace*{-1.5cm}-\,\frac{32}{3} \,\eta \,(1+\eta)\,(1-2\eta) \,\ln\eta - 
\frac{16}{3}\,(1-\eta)^2 \,(5+4\eta) \,\ln(1-\eta) 
\nonumber\\
&&\hspace*{-1.5cm}
+\, \frac{8}{3}\,(1-\eta)\,(5+9\eta-6\eta^2).\\[0.2cm]
g_2[^3\!P_1^{(1)}](\eta) &=& 
\frac{8}{3} \,(1-\eta)^2 
(1 + 2\eta) \left(3 \ln\frac{m_b^2}{\mu^2} + Y_R-X_R\right) 
\nonumber\\
&&\hspace*{-1.5cm}
+\, \frac{64}{3} \,\eta^2 
  \Bigg(-\mbox{Li}_2\Bigg(\frac{1-\eta}{2-\eta}\Bigg) + 
  \mbox{Li}_2\Bigg(\frac{2 (1-\eta)}{2-\eta}\Bigg) - 2 \mbox{Li}_2(\eta) - 
  \ln 2 \,\ln(2-\eta) + \ln 2 \,\ln\eta 
\nonumber\\
&&\hspace*{-1.5cm}
- \,\ln\eta \,\ln(1-\eta) + \frac{\pi^2}{3}\Bigg) + 
\frac{16 (1-\eta) \,(3+8 \eta^2-10 \eta^3+3 \eta^4)}{3 (2-\eta)^2}\, 
\ln(1-\eta) 
\nonumber\\
&&\hspace*{-1.5cm}
- \,\frac{8 \eta^2 \,(48-96 \eta+59 \eta^2-12 \eta^3)}{3 (2-\eta)^2}\, 
\ln\eta - 
\frac{32 (1-\eta) \,\eta \,(4-21 \eta+17 \eta^2-4 \eta^3)}{3 (2-\eta)^2}\,
 \ln 2 
\nonumber\\
&&\hspace*{-1.5cm}
- \,\frac{4 (1-\eta) \,(34+23 \eta-3 \eta^2-8 \eta^3)}{3 (2-\eta)}.\\[0.2cm]
g_3[^3\!P_1^{(1)}](\eta) &=& 
 \frac{16}{9}\,(1-\eta)^2 \,(1+2 \eta)\,\left[-\ln\frac{\hat{\mu}^2}{4 m_c^2}
+ 2 \ln(1-\eta)\right]\nonumber\\
&&\hspace*{-1.5cm}
-\,\frac{16}{9}\,(1-2\eta)\,(1+2\eta-2\eta^2)\,\ln\eta - 
 \frac{8}{27}\,(1-\eta)\,(13+37\eta-56\eta^2).
\end{eqnarray}

\subsubsection*{A.3.5. $^3\!P_2^{(1)}$}

\noindent
\begin{eqnarray}
g_1[^3\!P_2^{(1)}](\eta) &=& 0.\\[0.2cm]
g_2[^3\!P_2^{(1)}](\eta) &=& 0.\\[0.2cm]
g_3[^3\!P_2^{(1)}](\eta) &=& 
 \frac{16}{9}\,(1-\eta)^2 \,(1+2 \eta)\,\left[-\ln\frac{\hat{\mu}^2}{4 m_c^2}
+ 2 \ln(1-\eta)\right]\nonumber\\
&&\hspace*{-1.5cm}
-\,\frac{16}{45}\,(1-30\eta^2+20\eta^3)\,\ln\eta - 
 \frac{8}{45}\,(1-\eta)\,(27+31\eta-76\eta^2).
\end{eqnarray}

\subsubsection*{A.3.6. $^1\!P_1^{(1)}$}

\noindent
\begin{eqnarray}
g_1[^1\!P_1^{(1)}](\eta) &=& 0.\\[0.2cm]
g_2[^1\!P_1^{(1)}](\eta) &=& 0.\\[0.2cm]
g_3[^1\!P_1^{(1)}](\eta) &=& 
 \frac{16}{3}\,(1-\eta)^2 \,\left[-\ln\frac{\hat{\mu}^2}{4 m_c^2}
+ 2 \ln(1-\eta)\right]\nonumber\\
&&\hspace*{-1.5cm}
-\,\frac{8}{9}\,(1-6\eta+12\eta^2)\,\ln\eta - 
 \frac{4}{27}\,(1-\eta)\,(119-85\eta+8\eta^2).
\end{eqnarray}

\subsubsection*{A.3.7. $^3\!S_1^{(8)}$}

\noindent
\begin{eqnarray}
g_1[^3\!S_1^{(8)}](\eta) &=& 
-\frac{8\,(1-6\eta)}{3}\,\ln\eta+
 \frac{4}{9}\,(1 - \eta)\,(1 + 37\eta - 8\eta^2).\\[0.2cm]
g_2[^3\!S_1^{(8)}](\eta) &=& 
(1-\eta)^2 (1 + 2\eta) \left(3 \ln\frac{m_b^2}{\mu^2} + Y_R-X_R\right) 
\nonumber\\
&&\hspace*{-1.5cm}
+ \,\frac{2 (1-\eta)^3 (3-\eta^2)}{(2-\eta)^2} \,\ln(1-\eta) 
 -\Bigg(\frac{4 (1-\eta)^2 \eta^2}{2-\eta}+11\eta^2\Bigg) \,\ln\eta -
 \frac{8 (1-\eta)^3 \eta}{2-\eta} \,\ln 2 \nonumber\\
&&\hspace*{-1.5cm}
- \frac{(1-\eta) (34+23\eta-51\eta^2+16\eta^3)}{2 (2-\eta)}.\\[0.2cm]
g_3[^3\!S_1^{(8)}](\eta) &=& 
 \frac{3}{2} \,(1-\eta)^2 \,(1+2 \eta) \Bigg(-4 \ln\frac{m_b^2}{\mu^2} 
 + \frac{4}{3} X_R + \frac{14}{3} Y_R \nonumber\\
&&\hspace*{-1.5cm}
-\,\frac{2}{3} Z_R - 3 \ln^2(2-\eta) + 
 6 \ln(1-\eta) \,\ln(2-\eta)\Bigg) 
\nonumber\\
&&\hspace*{-1.5cm}
+ \,(1-\eta) \,(1+2 \eta) \Bigg((29+7 \eta)\,\mbox{Li}_2(\eta) - (7+29 \eta) 
 \,\frac{\pi^2}{6} + 9 \,(1+\eta) \,\mbox{Li}_2\Bigg(\frac{1-\eta}{2-\eta}
 \Bigg) 
\nonumber\\
&&\hspace*{-1.5cm}
- \,18 \,\mbox{Li}_2\Bigg(\frac{2 (1-\eta)}{2-\eta}\Bigg) + 18 \,\ln 2  
  \,\ln(2-\eta) 
- 18 \,\ln 2 \,\ln\eta + 2 \,(5+4 \eta) \,\ln\eta 
  \,\ln(1-\eta)\Bigg) 
\nonumber\\
&&\hspace*{-1.5cm}
-\,\frac{(1-\eta)^2 (110+188 \eta-291 \eta^2+83 \eta^3)}{(2-\eta)^2}\, 
  \ln(1-\eta) 
\nonumber\\
&&\hspace*{-1.5cm}
+\,\frac{34+547 \eta-1062 \eta^2+1158 \eta^3-426 \eta^4}{6 (2-\eta)}\,
 \ln\eta - \frac{(1-\eta)^2 (18+83 \eta-74 \eta^2)}{2-\eta}\,\ln 2 
\nonumber\\
&&\hspace*{-1.5cm}
+ \,\frac{(1-\eta)\,(5294+5779 \eta-14795 \eta^2+5228 \eta^3)}{36 (2-\eta)}.
\end{eqnarray}

\subsubsection*{A.3.8. $^1\!S_0^{(8)}$}

\noindent
\begin{eqnarray}
g_1[^1\!S_0^{(8)}](\eta) &=& 
 -8 \ln\eta-\frac{4}{3}\,(1-\eta) (11 - 7\eta + 2\eta^2).\\[0.2cm]
g_2[^1\!S_0^{(8)}](\eta) &=& 
3 (1-\eta)^2 \left(3 \ln\frac{m_b^2}{\mu^2} + Y_R-X_R\right) + 
\frac{6 \,(3-\eta) (1-\eta)^3}{(2-\eta)^2}\, \ln(1-\eta) 
\nonumber\\
&&\hspace*{-1.5cm}
+\,3\eta^2\ln\eta -\frac{3 (1-\eta) (34-53 \eta+17 \eta^2)}{2 (2-\eta)}.
\\[0.2cm]
g_3[^1\!S_0^{(8)}](\eta) &=& 
 \frac{9}{2} \,(1-\eta)^2 \Bigg(-4 \ln\frac{m_b^2}{\mu^2} + 
 \frac{4}{3} X_R + \frac{14}{3} Y_R - \frac{2}{3} Z_R 
\nonumber\\
&&\hspace*{-1.5cm}
 - \,3 \ln^2(2-\eta) + 
 6 \ln(1-\eta) \,\ln(2-\eta) - 6\,\ln 2\Bigg) 
\nonumber\\
&&\hspace*{-1.5cm}
+ \,3 \,(1-\eta) \Bigg((29+7 \eta)\,\mbox{Li}_2(\eta) - (7+29 \eta) 
 \,\frac{\pi^2}{6} + 9 \,(1+\eta) \,\mbox{Li}_2\Bigg(\frac{1-\eta}{2-\eta}
 \Bigg) 
\nonumber\\
&&\hspace*{-1.5cm}
- \,18 \,\mbox{Li}_2\Bigg(\frac{2 (1-\eta)}{2-\eta}\Bigg) + 18 \,\ln 2  
  \,\ln(2-\eta) 
- 18 \,\ln 2 \,\ln\eta + 2 \,(5+4 \eta) \,\ln\eta \,\ln(1-\eta)\Bigg) 
\nonumber\\
&&\hspace*{-1.5cm}
-\,\frac{3 (1-\eta)^2 (4+106 \eta-113 \eta^2+33 \eta^3)}{(2-\eta)^2\,\eta}\, 
  \ln(1-\eta) 
\nonumber\\
&&\hspace*{-1.5cm}
+\,\frac{17-48 \eta+90 \eta^2}{2}\,\ln\eta + 
 \frac{(1-\eta)\,(4478-6221 \eta+2077 \eta^2+20 \eta^3)}{12 (2-\eta)}.
\end{eqnarray}

\subsubsection*{A.3.9. $^3\!P_J^{(8)}$}

\noindent
We have summed over $J=0,1,2$.
\begin{eqnarray}
g_1[^3\!P_J^{(8)}](\eta) &=& 
 (1-\eta)^2 (1 + 2\eta) \Big[(-48)\,\ln\frac{\hat{\mu}^2}{4m_c^2}
 +96\,\ln(1-\eta)\Big]
\nonumber\\
&&\hspace*{-1.5cm}
 - \,24 \,(1 - 12\eta^2 + 8\eta^3) \ln\eta  
 - 4 (1 - \eta) \,(35 + 41\eta - 94\eta^2).\\[0.2cm]
g_2[^3\!P_J^{(8)}](\eta) &=& 
6 (1-\eta)^2 (1 + 2\eta) \left(3 \ln\frac{m_b^2}{\mu^2} + Y_R-X_R\right) 
\nonumber\\
&&\hspace*{-1.5cm}
+\, 48 \eta^2 \Bigg(-\mbox{Li}_2\Bigg(\frac{1-\eta}{2-\eta}\Bigg) + 
  \mbox{Li}_2\Bigg(\frac{2 (1-\eta)}{2-\eta}\Bigg) - 2 \mbox{Li}_2(\eta) - 
  \ln 2 \,\ln(2-\eta) + \ln 2 \,\ln\eta 
\nonumber\\
&&\hspace*{-1.5cm}
- \,\ln\eta \,\ln(1-\eta) + \frac{\pi^2}{3}\Bigg) + 
\frac{12 (1-\eta) \,(3+8 \eta^2-10 \eta^3+3 \eta^4)}{(2-\eta)^2}\, 
\ln(1-\eta) 
\nonumber\\
&&\hspace*{-1.5cm}
- \,\frac{6 \eta^2 \,(48-96 \eta+59 \eta^2-12 \eta^3)}{(2-\eta)^2}\, 
\ln\eta - 
\frac{24 (1-\eta) \,\eta \,(4-21 \eta+17 \eta^2-4 \eta^3)}{(2-\eta)^2}\,
 \ln 2 
\nonumber\\
&&\hspace*{-1.5cm}
- \,\frac{3 (1-\eta) \,(34+23 \eta-3 \eta^2-8 \eta^3)}{2-\eta}.\\[0.2cm]
g_3[^3\!P_J^{(8)}](\eta) &=& 
 (1-\eta)^2 \,(1+2 \eta)\,(-30)\,\ln\frac{\hat{\mu}^2}{4 m_c^2}
\nonumber\\
&&\hspace*{-1.5cm}
+\, 9 \,(1-\eta)^2 \,(1+2 \eta) \Bigg(-4 \ln\frac{m_b^2}{\mu^2} 
+ \frac{4}{3} X_R + \frac{14}{3} Y_R - \frac{2}{3} Z_R - 3 \ln^2(2-\eta) 
\nonumber\\
&&\hspace*{-1.5cm}
+ \,6 \ln(1-\eta) \,\ln(2-\eta)\Bigg) 
- 12 \,(3-4 \eta) \,(3+7 \eta) \Bigg(\mbox{Li}_2\Bigg(\frac{2 (1-\eta)}
 {2-\eta}\Bigg) - \ln 2 \,\ln(2-\eta) 
\nonumber\\
&&\hspace*{-1.5cm}
+ \,\ln 2 \,\ln\eta\Bigg) + 
 6 \,(29+36 \eta-91 \eta^2-14 \eta^3) \,\mbox{Li}_2(\eta) 
\nonumber\\
&&\hspace*{-1.5cm}
+ \,6 \,(9+18 \eta-29 \eta^2-18 \eta^3) \,\mbox{Li}_2\Bigg(\frac{1-\eta}
  {2-\eta}\Bigg) - 
 (7+36 \eta-25 \eta^2-58 \eta^3) \,\pi^2 
\nonumber\\
&&\hspace*{-1.5cm}
+ \,12 \,(5+9 \eta-16 \eta^2-8 \eta^3) \,\ln\eta \,\ln(1-\eta) 
\nonumber\\
&&\hspace*{-1.5cm}
- \,\frac{30 (1-\eta) (14+6 \eta-69 \eta^2+58 \eta^3-13 \eta^4)}{(2-\eta)^2}\,
  \ln(1-\eta) 
\nonumber\\
&&\hspace*{-1.5cm}
+ \,\frac{3 (4-68 \eta-159 \eta^2+632 \eta^3-466 \eta^4+106 \eta^5)}
  {(2-\eta)^2} \,\ln\eta 
\nonumber\\
&&\hspace*{-1.5cm}
- \,\frac{6 (1-\eta) (36+112 \eta-399 \eta^2+269 \eta^3-58 \eta^4)}
  {(2-\eta)^2}\, \ln 2 
\nonumber\\
&&\hspace*{-1.5cm}
+ \,\frac{(1-\eta) (1078+851 \eta-3757 \eta^2+1402 \eta^3)}{2 (2-\eta)}.
\end{eqnarray}

\subsubsection*{A.3.10. $^1\!P_1^{(8)}$}

\noindent
\begin{eqnarray}
g_1[^1\!P_1^{(8)}](\eta) &=&  
 16\,(1-\eta)^2 \,\left[-\ln\frac{\hat{\mu}^2}{4 m_c^2}
 + 2 \ln(1-\eta)\right]\nonumber\\
&&\hspace*{-1.5cm}
-\,\frac{8}{3}\,(1-6\eta+12\eta^2)\,\ln\eta - 
 \frac{4}{9}\,(1-\eta)\,(119-85\eta+8\eta^2).
\\[0.2cm]
g_2[^1\!P_1^{(8)}](\eta) &=& 0.\\[0.2cm]
g_3[^1\!P_1^{(8)}](\eta) &=& 
 10\,(1-\eta)^2 \,\left[-\ln\frac{\hat{\mu}^2}{4 m_c^2}
 + 2 \ln(1-\eta)\right]\nonumber\\
&&\hspace*{-1.5cm}
-\,\frac{2}{3}\,(7-15\eta+30\eta^2)\,\ln\eta - 
 \frac{1}{9}\,(1-\eta)\,(347-244\eta+29\eta^2).
\end{eqnarray}

\section*{Appendix B. Short-distance coefficients for energy 
distributions}
\setcounter{equation}{0}
\renewcommand{\theequation}{B.\arabic{equation}}

\noindent
At leading order in $\alpha_s$ the energy of the quarkonium is fixed. 
A non-trivial energy distribution is generated by gluon emission at 
NLO. The functions $g_i[n](\eta,x)$, where $x$ is the quarkonium 
energy fraction as defined in the text, can be expressed in the form 
\begin{equation}
\label{red}
g_i[n](\eta,x) = g_i[n](\eta)\,\delta(1+\eta-x) + 
[g_{i,real}[n](\eta,x)]_+,
\end{equation} 
where $g_i[n](\eta)$ is the corresponding function for the total 
integrated decay rate, given in Appendix A.3. The `plus-distribution' is 
defined by 
\begin{equation}
\int\limits_{2\sqrt{\eta}}^{1+\eta} d x\,\left[f(x)\right]_+\,t(x) 
= \int\limits_{2\sqrt{\eta}}^{1+\eta} d x\,f(x)\,(t(x)-t(1+\eta))
\end{equation}
for a test function $t$. We also introduce 
\begin{equation}
\rho=\sqrt{x^2-4\eta}.
\end{equation}
The kinematic limits on $x$ are $2\sqrt{\eta}<x<1+\eta$. In the 
following we give the energy distribution functions 
$g_{i,real}[n](\eta,x)$.

\subsection*{B.1. $^3\!S_1^{(1)}$}

\noindent
\begin{eqnarray}
g_{1,real}[^3\!S_1^{(1)}](x,\eta) &=& \frac{4}{3}
\Bigg(-\frac{7 (1+\eta-2 \eta^2)}{1+\eta-x} 
 -(3-2\eta)\Bigg)\rho 
\nonumber\\
&&\hspace*{-1.5cm}
+\frac{4}{3}\Bigg(\frac{4 (1-\eta)^2(1+2\eta)}{1+\eta-x}
 +2 (3+5\eta-2\eta^2-x-2\eta x)\Bigg) \ln\frac{
 2-x+\rho}{2-x-\rho}.\\[0.2cm]
g_{2,real}[^3\!S_1^{(1)}](x,\eta) &=& -\frac{8}{3} (1-2\eta)\rho 
+\frac{8}{3} (1-3\eta-x)\ln\frac{
2-x+\rho}{2-x-\rho}
\nonumber\\
&&\hspace*{-2cm}+\frac{32}{3}\, \eta^2 \ln\frac{
2\eta-x-\rho}{2\eta-x+\rho}.
\\[0.2cm]
g_{3,real}[^3\!S_1^{(1)}](x,\eta) &=& \frac{4}{3}\,(2-x)\rho 
+\frac{16}{3}\, x (1+\eta-x)\ln\frac{
2\eta-x-\rho}{2\eta-x+\rho}.
\end{eqnarray}

\subsection*{B.2. $^1\!S_0^{(1)}$}

\noindent
\begin{eqnarray}
g_{1,real}[^1\!S_0^{(1)}](x,\eta) &=&
\left(-\frac{28 (1-\eta)}{1+\eta-x} - 12\right)\,\rho
\nonumber\\
&&\hspace*{-1.5cm}
+\,\left(\frac{16\,(1-\eta)^2}{1+\eta-x} + 8 (3-\eta-x) \right)
\ln\frac{2-x+\rho}{2-x-\rho}.\\[0.2cm]
g_{2,real}[^1\!S_0^{(1)}](x,\eta) &=&-8 \,\rho
+ 8 (1+\eta-x)\, \ln\frac{2-x+\rho}{2-x-\rho}.\\[0.2cm]
g_{3,real}[^1\!S_0^{(1)}](x,\eta) &=&
(-4)\,(6+8\eta-7 x)\,\rho
+ 16\, (1+\eta-x)\,(x-2\eta) \,\ln\frac{2\eta-x-\rho}{2\eta-x+\rho}.
\end{eqnarray}

\subsection*{B.3. $^3\!P_0^{(1)}$}

\noindent
\begin{eqnarray}
g_{1,real}[^3\!P_0^{(1)}](x,\eta) &=& 0.\\[0.2cm]
g_{2,real}[^3\!P_0^{(1)}](x,\eta) &=& 0.\\[0.2cm]
g_{3,real}[^3\!P_0^{(1)}](x,\eta) &=&
\frac{8}{9}\left(\frac{4 (1-\eta)\,(1+2\eta)}{1+\eta-x} - \frac{1}{2}\,
 (26-40\eta+11 x)\right) \rho
\nonumber\\
&&\hspace*{-1.5cm}
+\frac{16}{3} \left(2 \eta^2+x-\eta x-x^2\right)
\ln\frac{2\eta-x-\rho}{2\eta-x+\rho}.
\end{eqnarray}

\subsection*{B.4. $^3\!P_1^{(1)}$}

\noindent
\begin{eqnarray}
g_{1,real}[^3\!P_1^{(1)}](x,\eta) &=&
\left(-\frac{8}{3}\right)
\left(\frac{7 (1-\eta) (1+2\eta)}{1+\eta-x} + 
3-2\eta \right)\,\rho
\nonumber\\
&&\hspace*{-1.5cm}
+\frac{16}{3}\left(\frac{2 (1-\eta)^2 (1+2\eta)}{1+\eta-x} + 
 3+5 \eta-2 \eta^2-x-2 \eta x \right)
\ln\frac{2-x+\rho}{2-x-\rho}.\\[0.2cm]
g_{2,real}[^3\!P_1^{(1)}](x,\eta) &=&
\frac{16}{3}\left(\frac{4\eta^2}{1+\eta-x} - (1-2\eta) \right)\,\rho
\nonumber\\
&&\hspace*{-1.5cm}
+\frac{16}{3}\left(\frac{4\eta^2}{1+\eta-x} + 1-3\eta-x \right)
\ln\frac{2-x+\rho}{2-x-\rho}
\nonumber\\
&&\hspace*{-1.5cm}
+\frac{32}{3}\left(-\frac{2\eta^2}{1+\eta-x} + 3\eta^2\right)
\ln\frac{2\eta-x-\rho}{2\eta-x+\rho}.\\[0.2cm]
g_{3,real}[^3\!P_1^{(1)}](x,\eta) &=&
\frac{8}{9}\left(\frac{4 (1-\eta) (1+2 \eta)}{1+\eta-x} - 
 (16+34 \eta-23 x)\right) \rho
\nonumber\\
&&\hspace*{-1.5cm}
-\frac{16}{3} \left(3 \eta+7 \eta^2-2 x-6 \eta x+2 x^2\right)
\ln\frac{2\eta-x-\rho}{2\eta-x+\rho}.
\end{eqnarray}

\subsection*{B.5. $^3\!P_2^{(1)}$}

\noindent
\begin{eqnarray}
g_{1,real}[^3\!P_2^{(1)}](x,\eta) &=& 0.\\[0.2cm]
g_{2,real}[^3\!P_2^{(1)}](x,\eta) &=& 0.\\[0.2cm]
g_{3,real}[^3\!P_2^{(1)}](x,\eta) &=&
\frac{8}{9} \left(\frac{4 (1-\eta) (1+2 \eta)}{1+\eta-x} - 
 (4-2 \eta+ x)\right)\rho
\nonumber\\
&&\hspace*{-1.5cm}
+ \,\frac{16}{15}\left(3 \eta-5 \eta^2+2 x-2 \eta x-2 x^2\right)
\ln\frac{2\eta-x-\rho}{2\eta-x+\rho}.
\end{eqnarray}

\subsection*{B.6. $^1\!P_1^{(1)}$}

\noindent
\begin{eqnarray}
g_{1,real}[^1\!P_1^{(1)}](x,\eta) &=& 0.\\[0.2cm]
g_{2,real}[^1\!P_1^{(1)}](x,\eta) &=& 0.\\[0.2cm]
g_{3,real}[^1\!P_1^{(1)}](x,\eta) &=&
\frac{4}{3}\left(\frac{8\,(1-\eta)}{1+\eta-x} - (14+x)\right)
\rho
\nonumber\\
&&\hspace*{-1.5cm}
-\,\frac{16}{3}\, (2 \eta-x-\eta x+x^2)\,
\ln\frac{2\eta-x-\rho}{2\eta-x+\rho}.
\end{eqnarray}

\subsection*{B.7. $^3\!S_1^{(8)}$}

\noindent
\begin{eqnarray}
g_{1,real}[^3\!S_1^{(8)}](x,\eta) &=&
4 (2-x)\,\rho
+ 16x (1+\eta-x)\,\ln\frac{2\eta-x-\rho}{2\eta-x+\rho}.\\[0.2cm]
g_{2,real}[^3\!S_1^{(8)}](x,\eta) &=&
-2 \left(1-2\eta\right)\,\rho
+2 \left(1 - 3\eta - x\right)
\ln\frac{2-x+\rho}{2-x-\rho}
\nonumber\\
&&\hspace*{-2cm}
+\,8\eta^2\,\ln\frac{2\eta-x-\rho}{2\eta-x+\rho}.\\[0.2cm]
g_{3,real}[^3\!S_1^{(8)}](x,\eta) &=&
\left(-\frac{32 (1-\eta) (1 + 2\eta)}{1+\eta-x} + 
 39 + 68\eta - \frac{77 x}{2}\right)\,\rho
\\
&&\hspace*{-3cm}
+\,\Bigg(-\frac{2 (1-\eta) (1 + 2\eta) (5 + 4\eta)}{1+\eta-x} + 
 17 + 5\eta - 8\eta^2 - 9 x - 8\eta x\Bigg)
\ln\frac{2-x+\rho}{2-x-\rho}
\nonumber\\
&&\hspace*{-3cm}
+\,\Bigg(\frac{18 (1 - \eta) (1 + 2\eta)}{1+\eta-x} 
 -18 + 18\eta + 92\eta^2 + 10 x - 
  26 \eta x - 10 x^2\Bigg)
\ln\frac{2\eta-x-\rho}{2\eta-x+\rho}.
\nonumber
\end{eqnarray}

\subsection*{B.8. $^1\!S_0^{(8)}$}

\noindent
\begin{eqnarray}
g_{1,real}[^1\!S_0^{(8)}](x,\eta) &=&
-12\left(6+8\eta-7 x\right)\,\rho
-48 (1+\eta-x) (2\eta-x)\,
\ln\frac{2\eta-x-\rho}{2\eta-x+\rho}.\\[0.2cm]
g_{2,real}[^1\!S_0^{(8)}](x,\eta) &=&
-6 \rho
+ 6 (1+\eta-x) \,\ln\frac{2-x+\rho}{2-x-\rho}.\\[0.2cm]
g_{3,real}[^1\!S_0^{(8)}](x,\eta) &=&
\left(-\frac{96 (1-\eta)}{1+\eta-x} -51-60\eta+\frac{105 x}{2}\right)\,\rho
\\
&&\hspace*{-1.5cm}
+\left(-\frac{6 (1-\eta) (5 + 4\eta)}{1+\eta-x} + 51 + 3\eta - 27 x\right)
\ln\frac{2-x+\rho}{2-x-\rho}
\nonumber\\
&&\hspace*{-1.5cm}
+\left(\frac{54 (1-\eta)}{1+\eta-x} -54 - 6\eta - 60\eta^2 + 30 x + 
  90\eta x - 30 x^2\right)
\ln\frac{2\eta-x-\rho}{2\eta-x+\rho}.
\nonumber
\end{eqnarray}

\subsection*{B.9. $^3\!P_J^{(8)}$}

\noindent
We have summed over $J=0,1,2$.
\begin{eqnarray}
g_{1,real}[^3\!P_J^{(8)}](x,\eta) &=&
\left(\frac{96 (1-\eta) (1+2\eta)}{1+\eta-x} - 
12 (18+16\eta-13 x) \right)\,\rho
\nonumber\\
&&\hspace*{-2cm}
+\,48 \left(-2\eta-8\eta^2+3 x+5 \eta x-3 x^2\right)
\ln\frac{2\eta-x-\rho}{2\eta-x+\rho}.\\[0.2cm]
g_{2,real}[^3\!P_J^{(8)}](x,\eta) &=&
\left(\frac{48\eta^2}{1+\eta-x} -12 (1-2\eta) \right)\,\rho
\nonumber\\
&&\hspace*{-2cm}
+\left(\frac{48\eta^2}{1+\eta-x} + 12 (1-3\eta-x) \right)
\ln\frac{2-x+\rho}{2-x-\rho}
\nonumber\\
&&\hspace*{-2cm}
+\left(-\frac{48\eta^2}{1+\eta-x} + 72\eta^2\right)
\ln\frac{2\eta-x-\rho}{2\eta-x+\rho}.\\[0.2cm]
g_{3,real}[^3\!P_J^{(8)}](x,\eta) &=&
\left(-\frac{12 (11+11\eta-32\eta^2)}{1+\eta-x} -24 (5-12\eta+x)\right)
\rho
\\
&&\hspace*{-3cm}
+\left(-\frac{12 (5+9\eta-16\eta^2-8\eta^3)}{1+\eta-x} + 
 6 (17-4\eta-8\eta^2-9 x-8\eta x\right)
\ln\frac{2-x+\rho}{2-x-\rho}
\nonumber\\
&&\hspace*{-3cm}
+\left(\frac{12 (3-4\eta) (3+7\eta)}{1+\eta-x} -
6 (18-8\eta-71\eta^2-24 x+2 \eta x+24 x^2) \right)
\ln\frac{2\eta-x-\rho}{2\eta-x+\rho}.
\nonumber
\end{eqnarray}

\subsection*{B.10. $^1\!P_1^{(8)}$}

\noindent
\begin{eqnarray}
g_{1,real}[^1\!P_1^{(8)}](x,\eta) &=&
\left(\frac{32 (1-\eta)}{1+\eta-x} - 
4 (14+x) \right)\,\rho
\nonumber\\
&&\hspace*{-1.5cm}
-\,16 \left(2 \eta-x-\eta x+x^2\right)
\ln\frac{2\eta-x-\rho}{2\eta-x+\rho}.\\[0.2cm]
g_{2,real}[^1\!P_1^{(8)}](x,\eta) &=& 0.\\[0.2cm]
g_{3,real}[^1\!P_1^{(8)}](x,\eta) &=&
\left(\frac{20 (1-\eta)}{1+\eta-x} - (62+36 \eta-29 x)\right)
\rho
\nonumber\\
&&\hspace*{-1.5cm}
-\,4\left(14 \eta+9 \eta^2-7 x-16 \eta x+7 x^2\right)
\ln\frac{2\eta-x-\rho}{2\eta-x+\rho}.
\end{eqnarray}

%%%%%%%%%%%%%%%%%%%%%%%%%%%%%%%%%%%%%%%%%%%%%%%%%%%%%%%%%%%%%%%%%%%%%%%

\end{document}